\def\lesssim{\mathrel{\hbox{\rlap{\hbox{\lower4pt\hbox{$\sim$}}}\hbox{$<$}}}}

\def\gtrsim{\mathrel{\hbox{\rlap{\hbox{\lower4pt\hbox{$\sim$}}}\hbox{$>$}}}}

\def\msun{M$_{\odot}$}

\def\teff{$T_{\rm eff}$}

\def\lteff{log $T_{\rm eff}$~}

\def\all_lsun{log$({L/\rm L_{\odot}})$~}

\def\masa_msun{$M/ \rm M_{\odot}$~}

\def\m_mstar{$M/M_{*}$~}

\def\mean#1{{\langle}#1{\rangle}}
\documentclass{aa}
\usepackage{graphicx}
        
\begin{document}

\title{New evolutionary models for massive ZZ Ceti stars: \\
I. First results for their pulsational properties}

\author{L. G. Althaus$^1$\thanks{Member of the Carrera del Investigador
Cient\'{\i}fico y Tecnol\'ogico, CONICET, Argentina.},
A. M. Serenelli$^1$\thanks{Fellow of CONICET, Argentina.},  
A. H. C\'orsico$^1$\thanks{Fellow of CONICET, Argentina.}, 
\and M. H. Montgomery$^2$ }
\offprints{L. G. Althaus}

\institute{
$^1$ Facultad  de  Ciencias
Astron\'omicas  y Geof\'{\i}sicas, Universidad  Nacional de  La Plata,
Paseo  del  Bosque S/N,  (1900)  La  Plata,  Argentina.\\ Instituto  de
Astrof\'{\i}sica de La Plata, IALP, CONICET\\
$^2$ Institute of Astronomy, Madingley Road, Cambridge CB3 0HA, UK\\
\email{althaus,serenell,acorsico@fcaglp.unlp.edu.ar,mikemon@ast.cam.ac.uk} }
\date{Received; accepted}

\abstract{
We   present   new   and   improved  evolutionary   calculations   for
carbon-oxygen  white dwarf  (WD) stars  appropriate for  the  study of
massive ZZ Ceti  stars. To this end, we  follow the complete evolution
of massive WD progenitors from  the zero-age main sequence through the
thermally pulsing  and mass  loss phases to  the WD  regime. Abundance
changes are accounted for by  means of a full coupling between nuclear
evolution  and  time-dependent  mixing  due  to  diffusive  overshoot,
semiconvection and  salt fingers. In  addition, time-dependent element
diffusion for  multicomponent gases has been considered  during the WD
stage. Emphasis is placed on  the chemistry variations along the whole
evolution. In  particular, we  find that before  the ZZ Ceti  stage is
reached,  element diffusion  has  strongly smoothed  out the  chemical
profile  to  such  a  degree  that the  resulting  internal  abundance
distribution does  not depend on the occurrence  of overshoot episodes
during the thermally pulsing phase.  The mass of the hydrogen envelope
left at the ZZ Ceti domain  amounts to $M_{\rm H} \approx$ 2.3 $\times
10^{-6}$ \msun.   This is  about half  as large as  for the  case when
element diffusion  is neglected.  The  implications of our  new models
for the pulsational properties of massive ZZ Ceti stars are discussed.
In  this regard,  we find  that  the occurrence  of core  overshooting
during  central   helium  burning   leaves  strong  imprints   on  the
theoretical  period spectrum of  massive ZZ  Ceti stars.   Finally, we
present  a simple new  prescription for  calculating the  He/H profile
which goes beyond the trace element approximation.
\keywords{stars:  evolution  --  stars: abundances -- stars: AGB
stars: interiors -- stars:white dwarfs -- stars: oscillations } }

\authorrunning{Althaus et al.}

\titlerunning{New evolutionary models for massive ZZ Ceti stars.}

\maketitle


\section{Introduction}

Variable DA white  dwarf (WD) stars or ZZ  Ceti stars belong currently
to one of  the best established and most  extensively studied class of
non-radial  pulsating stars  (see Gautschy  \& Saio  1995, 1996  for a
review).   These hydrogen-rich,  pulsating  WDs exhibit  multiperiodic
luminosity variations  caused by gravity-modes of  low harmonic degree
($\ell \leq  2$) and periods in  the range of 100-1200  s.  Such modes
were first thought to be  excited by the $\kappa$-mechanism (Winget et
al.  1981).  However, the dominance of convective  energy transport in
the driving region led Brickhill  (1991) and Goldreich \& Wu (1999) to
propose convective  driving as the  main driving mechanism  for g-mode
oscillations in the  DA WDs.  ZZ Ceti stars are  well known to pulsate
in  a narrow  effective temperature  (\teff) \  interval  ranging from
10700  K $\lesssim$  \teff $\lesssim  $ 12500  K.   Numerous important
studies have  been devoted to exploring the  pulsational properties of
ZZ Ceti  stars, amongst  them Tassoul et  al. (1990), Brassard  et al.
(1991; 1992ab), Gautschy et al. (1996) and Bradley (1996, 1998, 2001).

Over the last  past decade, the study  of the pulsational pattern
of variable  WDs through  asteroseismological techniques has  become a
very powerful tool for probing the internal structure and evolution of
these  stars.  Indeed,  asteroseismological  inferences have  provided
independent  valuable constraints  to fundamental  properties  such as
core composition, outer layer chemical stratification and stellar mass
(Pfeiffer et  al. 1996;  and Bradley 1998,  2001 amongst  others).  In
particular,  asteroseismology of  massive ZZ  Ceti stars  has recently
drawn  the attention  of researchers  in  view of  the possibility  it
offers  to place  constraints on  the crystallization  process  in the
interior of WDs (Montgomery \&  Winget 1999).  This has been motivated
by the  discovery of pulsations in  the star BPM 37093  (Kanaan et al.
1992), a  massive ZZ  Ceti star which  should be  largely crystallized
(Winget el al.  1997).

Needless to  say, a  detailed WD  modeling as well  as a  complete and
self-consistent treatment  of the evolutionary stages prior  to the WD
formation  are   required  for  an  adequate   interpretation  of  the
pulsational   patterns.    In   this   regard,  the   outer   chemical
stratification, the relics of the nucleosynthesis and mixing processes
which occurred along the  asymptotic giant branch (AGB) evolution (see
D'Antona \& Mazzitelli 1990), is  a matter of the utmost importance as
far  as   asteroseismology  of  ZZ  Ceti  stars   is  concerned.   The
construction of ZZ Ceti models based on full evolutionary calculations
has  recently been undertaken.   Indeed, Althaus  et al.   (2002) have
presented detailed models for a 0.563-\msun \ ZZ Ceti remnant based on
the complete  evolution of an initially  3 \msun \  star, using models
which take  into account the  chemical evolution during the  WD regime
caused by  time-dependent element  diffusion.  The exploration  of the
pulsational properties of such  models has been performed by C\'orsico
et al.   (2001; 2002), who have  found that mode  trapping effects are
considerably  weakened   as  a  result   of  the  smoothness   of  the
diffusion-modeled  chemical  profiles.   These studies  constitute  an
improvement  as  compared  with  most  of  the  existing  research  in
pulsating  DA WDs  which invokes  diffusive equilibrium  in  the trace
element approximation to assess  the shape of hydrogen-helium chemical
interface.

In this paper  we extend the calculations presented  in Althaus et al.
(2002) to the  case of  massive intermediate-mass stars.   As compared
with Althaus et al.  (2002) major improvements in the treatment of the
abundance  changes have  been  made.  In  particular,  we developed  a
time-dependent  scheme that  fully  couples abundance  changes due  to
nuclear      burning,      mixing      processes      and      element
diffusion. Time-dependent mixing due to semiconvection and salt finger
is  fully  taken  into  account  as  well  as  exponentially  decaying
diffusive overshooting  above and below {\it  any} formally convective
region.  In  particular, the presence  of some overshooting  below the
convective envelope during the thermal pulses has been shown by Herwig
et al.  (1997) to yield  third dredge-up and carbon-rich AGB stars for
relatively low initial mass progenitors  (see also Ventura et al. 1999
and Mazzitelli  et al.  1999).   In addition, the occurrence  of extra
mixing  below the  helium-flash convection  zone during  the thermally
pulsing  AGB phase  is supported  by recent  evolutionary calculations
(Herwig et  al.  1999 and Herwig  2000).  These studies  show that the
resulting  intershell  abundances  are  in  agreement  with  abundance
determinations in hydrogen-deficient post-AGB remnants such as PG 1159
stars.

The main aim of this work  is to present the first results of detailed
and {\it complete evolutionary calculations} appropriate for the study
of massive  ZZ Ceti  stars. To  this end, we  follow the  evolution of
initially 7.5-  and 6-\msun  \ stars from  the zero-age  main sequence
through  the   thermally  pulsing   phase  on  the   AGB  to   the  WD
regime. Attention  is concentrated  on the chemistry  variations along
the  whole  evolution.   We   emphasize  in  particular  the  role  of
time-dependent   element   diffusion   on   the   chemical   abundance
distribution at the ZZ Ceti  stage. The exploration of the pulsational
properties of massive ZZ Ceti stars  in the light of our new models is
likewise within the scope of  this investigation. Next, in Sect. 2, we
briefly describe the main  physical inputs of the models, particularly
our treatment  for the chemical evolution and  overshooting.  In Sect.
3 we  present the main  evolutionary results. Pulsational  results are
discussed in  Sect. 4. In that section we also present a simple new 
prescription for calculating the He/H profile, which constitutes an 
improvement over the trace element approximation. Finally, Sect.  5  
is devoted to  making some
concluding remarks.

\section{Input of the models and evolutionary sequences}

Here  we describe  at  some  length the  main  characteristics of  our
evolutionary code.  We restrict ourselves  to the main updates  in the
macrophysics,  particularly the  treatment of  the  chemical abundance
changes. \\

\noindent{\sl General description of the code:}
The results presented in this work have been obtained with the stellar
evolution code LPCODE we employed  in our previous works. The code has
been developed at La Plata  Observatory and it is described in 
Althaus et  al. (2002) and  references therein.
For the purposes of the present paper, the code has been substantially
modified particularly  with regard to  the treatment of  the abundance
changes, modifications which will be described later in this section.

Briefly, the code  is based on the method of  Kippenhahn et al. (1967)
for  calculating   stellar  evolution.   Envelope   integrations  from
photospheric starting  values inward to a fitting  outer mass fraction
(close to the photosphere) are performed to specify the outer boundary
conditions.  The  independent variable is $\xi$=  ln $(1-M_r/M_*)$ and
the dependent variables are:  radius ($r$), pressure ($P$), luminosity
($l$)  and temperature ($T$).   The following  change of  variables is
considered in LPCODE:

\begin{eqnarray}
\theta^{(n+1)} &=& \theta^{(n)} + \ln{\left(  1  +  u_\theta \right)} 
\nonumber \\
p^{(n+1)}  &=& p^{(n)}  + \ln{\left(  1  + u_p  \right)} \nonumber  \\
x^{(n+1)}  &=& x^{(n)}  + \ln{\left(  1  + u_x  \right)} \nonumber  \\
l^{(n+1)} &=& l^{(n)} + u_l
\end{eqnarray}

\noindent with $u_\theta$, $u_p$, $u_x$ and $u_l$ being the quantities to be 
iterated  that  are   given  by  $u_\theta=  \displaystyle\frac{\Delta
T}{T^{(n)}}$,  $u_p=   \displaystyle\frac{\Delta  P}{P^{(n)}}$,  $u_x=
\displaystyle\frac{\Delta  r}{r^{(n)}}$ and  $u_l= \displaystyle\Delta
l$, where superscripts  $n$ and $n+1$ denote the  beginning and end of
time  interval (see Kippenhahn  et al.  1967 for  definitions).  Here,
$\theta=$  ln $T$,  $x=$ ln  $r$ and  $p=$ ln  $P$.  Thus,  the Henyey
iteration  scheme is  applied to  the differences  in  the luminosity,
pressure, temperature and radius between the previous and the computed
model.

As  for  the  constitutive  physics,  LPCODE  employs  OPAL  radiative
opacities  (including   carbon-  and  oxygen-rich   compositions)  for
arbitrary  metallicity  from  Iglesias   \&  Rogers  (1996)  and  from
Alexander  \&  Ferguson (1994)  for  the  low-temperature regime.   In
particular,  opacities for various  metallicities are  required during
the  WD  cooling regime  in  view  of  the metallicity  gradient  that
develops in  the envelope of the  models as a  result of gravitational
settling.  The equation of  state for the low-density regime comprises
partial  ionization for  hydrogen and  helium  compositions, radiation
pressure  and  ionic  contributions.   For  the  high-density  regime,
partially  degenerate  electrons  and  Coulomb interactions  are  also
considered.  For the  WD regime, we include an  updated version of the
equation of  state of Magni  \& Mazzitelli (1979). In LPCODE 
crystallization sets in when the ion coupling constant 
$\Gamma \equiv  Z^2 e^2/\mean{r}  k_B T$ reaches the value 180.
Neutrino emission
rates and  conductive opacities are taken  from the works  of Itoh and
collaborators (see  Althaus et  al.  2002).  A  nuclear network  of 34
thermonuclear reaction  rates and 16  isotopes has been  considered to
describe  the  hydrogen (proton-proton  chain  and  CNO bi-cycle)  and
helium burning, and carbon ignition.  Nuclear reaction rates are taken
from Caughlan \& Fowler  (1988), except for the reactions $^{15}$N($p,
\gamma)^{16}$O,      $^{15}$N($p,     \alpha)^{12}$C,     $^{18}$O($p,
\alpha)^{15}$N,              $^{18}$O($p,              \gamma)^{19}$F,
$^{12}$C($\alpha,\gamma)^{16}$O,      $^{16}$O($\alpha,\gamma)^{20}$Ne,
$^{13}$C($\alpha,n)^{16}$O,           $^{18}$O($\alpha,\gamma)^{22}$Ne,
$^{22}$Ne($\alpha,n)^{25}$Mg   and  $^{22}$Ne($\alpha,\gamma)^{26}$Mg,
which  are  taken from  Angulo  et  al.  (1999).  In  particular,  the
$^{12}$C($\alpha,\gamma)^{16}$O  reaction  rate  given  by  Angulo  et
al.  (1999) is  about twice  as large  as that  of Caughlan  \& Fowler
(1988).

To get a reasonable numerical accuracy, AGB models typically contained
1400  mesh points, except  for the  peak of  the thermal  pulses where
about 2000  mesh points were required. Evolutionary  time steps during
the thermally pulsing  phase ranged from a few  days during the helium
flashes and the subsequent phases  where the third dredge-up may occur
to some years during the stationary hydrogen-burning interpulse phase.
Finally, mesh distribution is performed every three time steps. \\

\noindent{\sl Chemical evolution:} 
An  important aspect  of  the present  study  is the  modeling of  the
chemical  abundance  distribution  throughout  all  of  the  different
evolutionary phases.  To this end, we consider a time-dependent scheme
for the  simultaneous treatment of chemical changes  caused by nuclear
burning  and  convective, semiconvective,  salt  finger and  overshoot
mixing. Needless to say, such a coupling between nuclear evolution and
time-dependent  mixing   is  much  more  physically   sound  than  the
instantaneous   mixing  approximation   usually  assumed   in  stellar
modeling.  In particular,  this kind  of  treatment has  been used  by
Mazzitelli  et al.   (1999) to  study  the lithium  production by  hot
bottom burning in AGB stars (see also Ventura et al. 1999).

In what  follows, we present  some details about the  numerical method
for  the  abundance changes  included  in  LPCODE.  Specifically,  the
abundance changes for  all chemical elements are described  by the set
of equations

\begin{equation}
\left( \frac{d \vec{Y}}{dt} \right) = 
\left( \frac{\partial \vec{Y}}{\partial t} \right)_{\rm nuc} +
\frac{\partial}{\partial M_r} \left[ (4\pi r^2 \rho)^2 D 
\frac{\partial \vec{Y}}{\partial M_r}\right], 
\end{equation}

\noindent with $\vec{Y}$ being the vector containing the number 
fraction  of all  considered  nuclear species.   Here,  mixing due  to
convection, semiconvection, salt finger  and overshoot is treated as a
diffusion process which is described by the second term of Eq.  (2) in
terms  of  the  mass  coordinate  $M_r$.   The  efficiency  of  mixing
processes is described by  appropriate diffusion coefficients $D$ (see
later in this section).  The first term of Eq. (2) gives the abundance
changes due to thermonuclear  reactions, changes which, in the present
work, are fully coupled to mixing processes.  This term is expanded as
a  function  of local  abundances  and  cross  sections following  the
implicit scheme  by Arnett \&  Truran (1969). After  linearization, it
can be written as

\begin{equation}
\left( \frac{\vec{Y}_j^{n+1} - \vec{Y}_j^n}{\Delta t} \right)_{\rm nuc} =
- \ \Gamma_j \ \vec{Y}^{n+1} + \vec{\Lambda}_j ,
\end{equation}

\noindent where $\Delta t$  is the time interval $\Delta t= t^{n+1} - t^{n}$
and superscripts $n$ and $n+1$  again denotes the beginning and end of
time  interval. Subscript  $j$ denotes  the $j^{\rm  th}$  mesh point.
$\Gamma_j$ is a  $N \times N$ matrix ($N$ being  the number of nuclear
species considered)  and $\Lambda_j$ a  vector of $N$  dimension, both
with  elements   involving  abundances  ($\vec{Y}_j^n$)   and  nuclear
reaction  rates at  $t^{n}$.  Using  a  three-point, finite-difference
scheme, the second term of Eq. (2) can be approximated by

$$
\hspace{-3.75cm}
\left\{ \frac{\partial}{\partial M_r} \left[ (4\pi r^2 \rho)^2 D 
\frac{\partial \vec{Y}}{\partial M_r}\right] \right\}_j^{n+1}=
$$
\begin{equation}
\hspace{3.7cm}
{\cal A} \vec{Y}_{j-1}^{n+1} + {\cal B} \vec{Y}_j^{n+1} + {\cal C} 
\vec{Y}_{j+1}^{n+1}.
\end{equation}
\indent Here
${\cal A}$, ${\cal B}$ and ${\cal C}$ are diagonal matrices with dimension
$N \times N$. Specifically, 
the non-vanishing elements of such matrices are given by

\begin{equation}
a_{ii} = \frac{ (4 \pi)^2 (\rho^2 r^4 D)_{j-1/2}}{(m_{j-1/2}-m_{j+1/2})
(m_{j-1}-m_j)} 
\end{equation}

\begin{equation}
c_{ii} = \frac{ (4 \pi)^2 (\rho^2 r^4 D)_{j+1/2}}{(m_{j-1/2}-m_{j+1/2})
(m_{j}-m_{j+1})} 
\end{equation}

\begin{equation}
b_{ii} = - ( a_{ii} + c_{ii} ),
\end{equation}

\noindent where the subscripts $j+1/2$ ($j-1/2$) indicate an average for the
mass shell between the mesh-points $j$ and $j+1$ ($j-1$). At the boundaries of 
the mass interval of integration, we have no mass flux 
($\partial \vec{Y} / \partial M_r=0$), thus
\begin{equation}
\frac{\partial}{\partial M_r} \left[ (4\pi r^2 \rho)^2 D 
\frac{\partial \vec{Y}}{\partial M_r}\right]_{\rm bdry} = 
(4\pi r^2 \rho)^2 D 
\frac{\partial^2 \vec{Y}}{\partial M_r^2}.
\end{equation}

A second-order Taylor expansion in the abundances leads us to

\begin{equation}
c_{ii} = 2 \frac{ (4 \pi)^2 (\rho^2 r^4 D)_{1/2}}{(m_{1}-m_{2})^2}; \ \ \ 
b_{ii} = - c_{ii}; \ \ \ a_{ii}= 0
\end{equation} 

\noindent at the top boundary, and to

\begin{equation}
a_{ii} = 2 \frac{ (4 \pi)^2 (\rho^2 r^4 D)_{J-1/2}}{(m_{J-1}-m_{J})^2}; \ \ \ 
b_{ii} = - a_{ii}; \ \ \ c_{ii}= 0
\end{equation}

\noindent at the bottom boundary (characterized by the $J^{\rm th}$ 
mesh-point).  

Eqs. (2), (3) and (4) lead to the following system of linear equations
to  be   solved  simultaneously   for  the  new   chemical  abundances
$\vec{Y}^{n+1}$ at time $t^{n+1}$:

\begin{equation}
{\cal F}_j  \vec{Y}^{n+1}_j -  {\cal A}_j \vec{Y}^{n+1}_{j-1}  - {\cal
C}_j \vec{Y}^{n+1}_{j+1} =
\vec{\cal M}_j
\end{equation}

\noindent where ${\cal F}_j = {\cal T}^{-1}+\Gamma_j-{\cal B}_j$ and 
$\vec{\cal M}_j=  \vec{\Lambda}_j +  \vec{Y}^n_j / \Delta  t$.  ${\cal
T}^{-1}$ is a $N \times N$ diagonal matrix with elements\footnote{Note
that  Eq. (11) couples  nuclear evolution  to the  current composition
change due to mixing  processes.} $1/\Delta t$.  Fortunately, Eq. (11)
has  a special structure  (tridiagonal) which  enables us  to minimize
storage of matrix coefficients.  Schematically, we have (${\cal A}_j$,
${\cal F}_j$  and ${\cal C}_j$ are matrices  and $\vec{Y}_j^{n+1}$ and
$\vec{\cal M}_j$ are vectors)

$$
\left( 
\begin{array}{cccccccccc}
{\cal F}_1 & {\cal C}_2 & & & & & & & & \\
{\cal A}_1 & {\cal F}_2 & {\cal C}_3 & & & & & & & \\
& {\cal A}_2 & {\cal F}_3 & {\cal C}_4 & & & & & & \\
& & \ddots & \ddots & \ddots & & & & & \\
& & & \ddots & \ddots & \ddots & & & & \\
& & & & \ddots & \ddots & \ddots & & &  \\
& & & & & {\cal A}_{J-3} & {\cal F}_{J-2} & {\cal C}_{J-1} &  \\
& & & & & & {\cal A}_{J-2} & {\cal F}_{J-1} & {\cal C}_{J}  \\
& & & & & & & {\cal A}_{J-1} & {\cal F}_{J} 
\end{array}
\right)
\bullet 
$$
\begin{equation}
\hspace{2cm}
\bullet
\left( 
\begin{array}{c}
\vec{Y}_1^{n+1} \\
\vec{Y}_2^{n+1} \\
\vec{Y}_3^{n+1} \\
\vdots \\
\vdots \\
\vdots \\
\vec{Y}_{J-2}^{n+1}\\
\vec{Y}_{J-1}^{n+1}\\
\vec{Y}_{J}^{n+1}
\end{array}
\right)
= 
\left( 
\begin{array}{c}
{\vec{\cal M}}_1 \\
{\vec{\cal M}}_2 \\
{\vec{\cal M}}_3 \\
\vdots \\
\vdots \\
\vdots \\
{\vec{\cal M}}_{J-2}\\
{\vec{\cal M}}_{J-1}\\
{\vec{\cal M}}_{J}
\end{array}
\right)
\end{equation}

To solve this  set of equations, we followed  the method considered by
Iben \&  MacDonald (1985).  Eq.   (2) is applied to  radiative ($D=0$)
zones  and to  convective, semiconvective,  overshoot and  salt finger
regions  provided  the  diffusion  coefficient  $D$  is  appropriately
specified  for each  process  according to  the  adopted treatment  of
convection.   We consider a  total of  16 isotopes  ($N=16$): $^{1}$H,
$^{2}$H, $^{3}$He,  $^{4}$He, $^{7}$Li, $^{7}$Be,  $^{12}$C, $^{13}$C,
$^{14}$N, $^{15}$N, $^{16}$O,  $^{17}$O, $^{18}$O, $^{19}$F, $^{20}$Ne
and  $^{22}$Ne.   In LPCODE,  abundance  changes  are performed  after
convergence of each  stellar model (and not during  iterations). For a
better  integration precision  in the  nuclear  evolution computation,
each evolutionary time step is divided into 5 chemical time steps.

The evolution of the chemical abundance distribution caused by element
diffusion during the whole WD evolution constitutes an important point
of the present work.  In  our treatment of time-dependent diffusion we
have  considered  gravitational  settling  and  chemical  and  thermal
diffusion  for  the  following  nuclear  species:  $^{1}$H,  $^{3}$He,
$^{4}$He,  $^{12}$C,  $^{14}$N and  $^{16}$O.  The chemical  evolution
resulting  from element diffusion  is described,  for a  given isotope
$i$, by the continuity equation as

\begin{equation}
\left( \frac{\partial Y_i}{\partial t} \right)_{\rm diff}=
-\frac{\partial}{\partial M_r} \left( 4\pi r^2 \rho Y_i w_i
\right), 
\end{equation}

\noindent where $w_i$ is the diffusion velocity. We have adopted the 
treatment of Burgers (1969) for multicomponent gases.   
In  this work,  our  focus  is  on the  chemical  evolution
occurring quite deep in the  star, thus radiative levitation (which is
expected  to modify  the  surface  composition of  hot  WDs) has  been
neglected.  In  terms of  the  gradient  of  ion densities,  diffusion
velocities can be written in the form

\begin{equation}
w_i = w_{i}^{\rm gt} - \sum\limits_{{\rm ions} (j)} \sigma_{ij}\
\frac{d\ln{n_j}}{dr} ,
\end{equation}

\noindent where $w_{i}^{\rm gt}$ stands for the velocity component due to
gravitational  settling and  thermal diffusion.  Details are  given in
Althaus \& Benvenuto  (2000); see also Gautschy \&  Althaus (2002) for
an application  to DB WDs.   Unlike our previous works,  the abundance
changes resulting from element  diffusion are fully coupled to nuclear
reactions.  To this  end, matrix  coefficients in  Eq. (12)  have been
appropriately modified. \\

\noindent{\sl Overshooting:}\ 
In the present study  we have included time-dependent overshoot mixing
during all  pre-WD evolutionary stages.  The scheme  for the abundance
changes  described above  enables  us a  self-consistent treatment  of
diffusive  overshooting  in  the  presence  of  nuclear  burning.   In
particular,  we  have   considered  exponentially  decaying  diffusive
overshooting  above and  below {\it  any} formally  convective region,
including  the  convective  core  (main sequence  and  central  helium
burning phases), the external  convective envelope and the short-lived
helium-flash convection zone which develops during the thermal pulses.
Specifically, we have followed the formalism of Herwig (2000) based on
the hydrodynamical  simulations of Freytag et al.   (1996), which show
that turbulent  velocities decay exponentially  outside the convective
boundaries.  The expression for the diffusion coefficient in overshoot
regions is
 
\begin{equation}
D_{\rm os} = D_0 \ {\rm exp} \left(\frac{-2 \ z}{ H_{\rm v}} \right),
\end{equation}

\noindent where $D_0$ is the diffusion coefficient at the boundary of
the convection zone  (see below), $z$ is the  radial distance from the
edge of the convection zone, $ H_{\rm v}= f H_{\rm P}$, where the free
parameter $f$ is a measure of  the extent of the overshoot region, and
$H_{\rm P}$ is  the pressure scale height at  the convective boundary.
In  this study  we have  assumed $f=  0.015$, which  accounts  for the
observed  width  of  the  main  sequence as  well  as  the  intershell
abundances of hydrogen-deficient post-AGB  remnants (see Herwig et al.
1997;  1999,   Herwig  2000  for  details;  see   also  Mazzitelli  et
al. 1999).\\

\noindent{\sl Treatment of convection:} For this work, we included in
LPCODE the extended mixing length theory of convection for fluids with
composition gradients developed by Grossman et al. (1993) in its local
approximation as given by Grossman \& Taam (1996).  These authors have
developed  the non-linear  mixing  length theory  of double  diffusive
convection that applies in  convective, semiconvective and salt finger
instability  regimes.   According  to  this treatment,  the  diffusion
coefficient  $D$ in  Eq.  (2)  characterizing such  mixing  regimes is
given by

\begin{equation}
D= \ell \sigma
\end{equation}

\noindent where $\ell=  \alpha H_P$ is  the mixing length and $\sigma$ the  
turbulent   velocity.  The   value  of   $\sigma$  is   determined  by
simultaneously solving  the equations  for the turbulent  velocity and
flux conservation (Eqs.  9 and 17 of Grossman and  Taam 1996). In this
work, the free parameter $\alpha$ is taken to be 1.5. \\

\noindent{\sl Mass loss:} Our treatment  of mass loss  is that 
of Bl\"ocker (1995). In particular, during the AGB evolution, the mass
loss rate is given by

\begin{equation}
\dot M= 4.83 \times 10^{-9} M_{\rm ZAMS}^{-2.1} L^{2.7} \dot M_{\rm R} \ 
[M_{\odot}/yr]
\end{equation}

\noindent where $ \dot M_{\rm R} $ is the usual Reimer rate given by
$\dot M_{\rm R}=  4 \times 10^{-13} \eta_{\rm R} L  R/M$ with $L$, $R$
and $M$  the luminosity, radius and  mass of the star  in solar units.
This formulation,  based on dynamical computations  for atmospheres of
Mira-like stars,  accounts for  the strong increase  of the  mass loss
rate and modulations  by thermal pulses expected in  the course of the
AGB evolution.  We set the  value of the free parameter $\eta_{\rm R}$
somewhat  arbitrarily at  1 in  order to  get a  reasonable  number of
thermal pulses  on the AGB.   A more careful  choice for the  value of
$\eta_{\rm R}$ is not relevant  for the purposes of the present paper.
In our computation, mass loss  episodes taking place during the stages
of core helium  burning and red giant branch  are considered according
to the usual Reimer's formulation with $\eta_{\rm R}$=1.\\

\noindent{\sl Global properties of the evolutionary computations:} 
A primary target of this  work is the construction of detailed massive
DA  WD  models  with   $^{12}$C/$^{16}$O  cores  appropriate  for  the
pulsational studies.  Because  the self-consistent solution of nuclear
evolution and time-dependent mixing demands a considerable increase of
computing  time,   particularly  during  the  AGB  phase   of  the  WD
progenitor, we restrict ourselves  to examining two cases of evolution
for the progenitor: \\

\begin{itemize}

\item As for the first case, which will be hereinafter referred to as 
sequence NOV, we computed the {\it complete} evolution of an initially
7.5-\msun \  stellar model  {\it without} overshooting.   The sequence
was  followed from the  zero-age main  sequence all  the way  from the
stages of hydrogen and helium burning  in the core up to the thermally
pulsing  phase at  the tip  of  the AGB.   After experiencing  several
thermal pulses and  as a result of mass  loss episodes, the progenitor
departs    from    the    AGB    and   evolves    towards    its    WD
configuration. Specifically,  during the  7th pulse, an  enhanced mass
loss rate is invoked to terminate the AGB phase. Evolution was pursued
to the  domain of  the ZZ Ceti  stars on  the WD cooling  branch.  The
final mass of the WD remnant is about 0.936 \msun. \\

\item As for the second case (sequence OV), we followed the
evolution of a 6-\msun \  progenitor from the main sequences until the
star underwent four thermal  pulses.  Here, diffusive overshooting was
applied  during   all  evolutionary  stages  and   to  all  convective
boundaries.   This sequence  was done  with the  aim of  assessing the
influence of  overshooting on the WD  internal chemical stratification
and  to explore  its implications  for the  pulsational  properties of
massive ZZ Ceti stars as well. For this sequence, we have not computed
the   post-AGB  evolution.    Rather,  we   obtained  an   initial  WD
configuration by  simply scaling the chemical  profile built-up during
the pre-WD evolution to the structure of a hot WD model resulting from
sequence  NOV.   Because  we  are  interested in  the  final  chemical
stratification of  ZZ Ceti  models, this procedure  is enough  for our
purposes. Here, the mass of the $^{12}$C/$^{16}$O core towards the end
of thermal pulses amounts to 0.94 \msun.

\end{itemize}

For   both  sequences,   a  solar-like   initial   composition  (Y,Z)=
(0.275,0.02)  corresponding  to Anders  \&  Grevesse  (1989) has  been
adopted.  Interestingly,  the mass of  the resulting $^{12}$C/$^{16}$O
core is  quite similar  in both sequences  despite the  very different
initial  mass  values  of  the  progenitors (see  also  Mazzitelli  et
al.  1999). In  fact, overshooting  leads to  larger  convective cores
during the core burning stages.  We want to mention that the breathing
pulse instability occurring towards the end of core helium burning has
been suppressed  (see Straniero  et al. 2003  for a  recent discussion
about  this  point).   Finally,  for  both  sequences,  time-dependent
element diffusion was considered during the whole WD evolution.

To the  best of our knowledge, this  is the first time  that WD models
appropriate for the study of pulsational properties of massive ZZ Ceti
stars  are  derived  from  detailed  evolutionary  calculations  which
include   a  self-consistent   treatment  of   time-dependent  element
diffusion and  nuclear burning.   We report below the predictions of
our  calculations, particularly  for the  chemical  stratification and
comment on their implications for the relevant pulsation properties.

\section{Evolutionary results}

\subsection{Evolution of the white dwarf progenitor}

In this  section, we  describe the main  results for  the evolutionary
stages prior  to the  WD formation.  Attention  will be  restricted to
analysing the relevant aspects for the WD formation, in particular the
chemistry variations along the  evolution.  Although the full coupling
between  nuclear  evolution  and   mixing  implemented  in  LPCODE  is
appropriate  for addressing problems  such as  hot bottom  burning and
lithium  production in AGB  stars, we  will not  explore them  in this
paper.  Such specific issues  would carry  us too  far afield,  and we
refer the  reader to  the works of  D'Antona \& Mazzitelli  (1996) and
Mazzitelli et al.  (1999)  for details.  Other recent relevant studies
of  AGB   evolution  are  those  of  Sackmann   \&  Boothroyd  (1992),
Vassiliadis  \&  Wood  (1993),  Bl\"ocker  (1995),  Straniero  et  al.
(1997),  Wagenhuber \&  Groenewegen (1998)  and Herwig  (2000) amongst
others.

\begin{figure}
\centering
\includegraphics[clip,width=250pt]{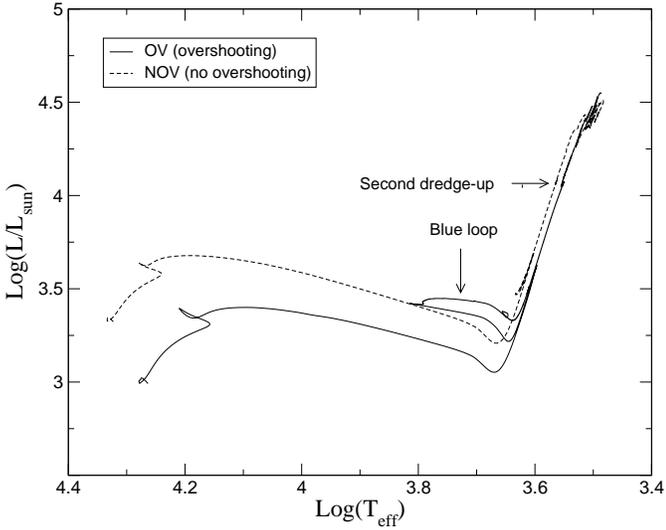}
\caption{The Hertzsprung-Russell diagram for the evolution of our
7.5- and 6-M$_{\odot}$ sequences (NOV and OV, respectively) from
the main sequence to the first thermal pulses.}
\end{figure}

We begin by examining Fig. 1 which illustrates the Hertzsprung-Russell
(HR)  diagram for  the  WD progenitor  from  the main  sequence to  an
advanced AGB phase. Solid and  dashed lines correspond to sequences OV
and  NOV, respectively.   A  feature worthy  of  comment predicted  by
sequence OV is that central helium burning occurs mostly during a loop
to the blue in the HR diagram.  Indeed, the blue excursion begins when
the central  helium abundance by  mass falls below $\approx$  0.75 and
continues  until the  abundance  has decreased  below  0.1.  For  this
sequence, the total  time spent during hydrogen and  helium burning in
the core  amounts to  $7.5 \times 10^{7}$  yr, while for  sequence NOV
this  time is  $4.2 \times  10^{7}$ yr.   Following the  exhaustion of
central  helium and  before the  re-ignition of  the  hydrogen burning
shell {\it both of our sequences experience the second dredge-up}.  As
a   result,   the  surface   composition   is  appreciably   modified,
particularly  for  sequence  OV.   In  fact, for  this  sequence,  the
envelope helium  abundance rises from  about 0.30 (resulting  from the
first  dredge-up) to 0.364.   For sequence  NOV, the  second dredge-up
brings the helium abundance to 0.328. It is also worth mentioning that
during the second dredge-up  phase, the mass of the hydrogen-exhausted
core is  strongly reduced. In  particular, for sequence NOV,  the core
mass is  reduced from  $\approx$ 1.50  \msun \ at  the end  of central
helium  burning  to  $\approx$   0.93  \msun  \  after  the  dredge-up
episode. For sequence OV,  the hydrogen-exhausted core is reduced from
1.38 to 0.938 \msun \ during the dredge-up.

\begin{figure}
\centering
\includegraphics[clip,width=250pt]{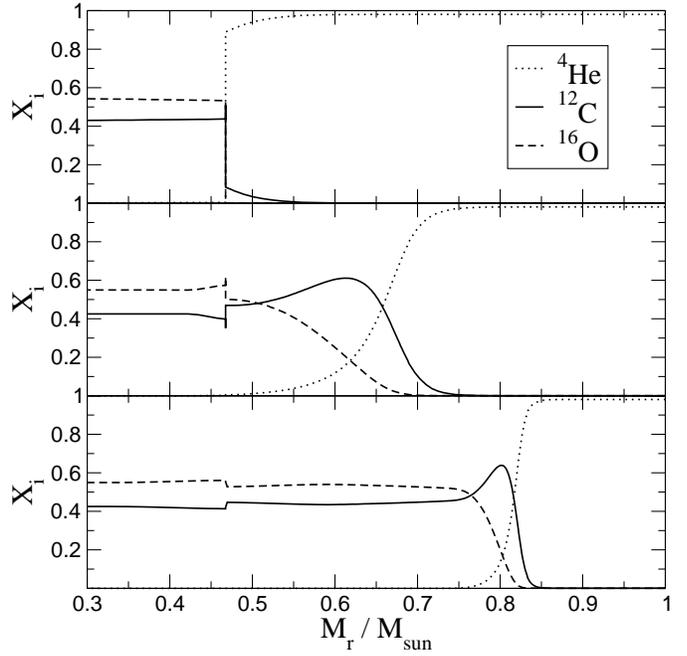}
\caption{Internal $^{4}$He,  $^{12}$C and $^{16}$O profiles for the
7.5-M$_{\odot}$ sequence (NOV) at  three evolutionary stages after the end
of core  helium burning.  A  salt finger instability develops  at $M_r
\approx$  0.45  M$_{\odot}$  (middle  panel)  which  leads  to  a
chemical redistribution there
(bottom panel). See text for details.}
\end{figure}

\begin{figure}
\centering
\includegraphics[clip,width=250pt]{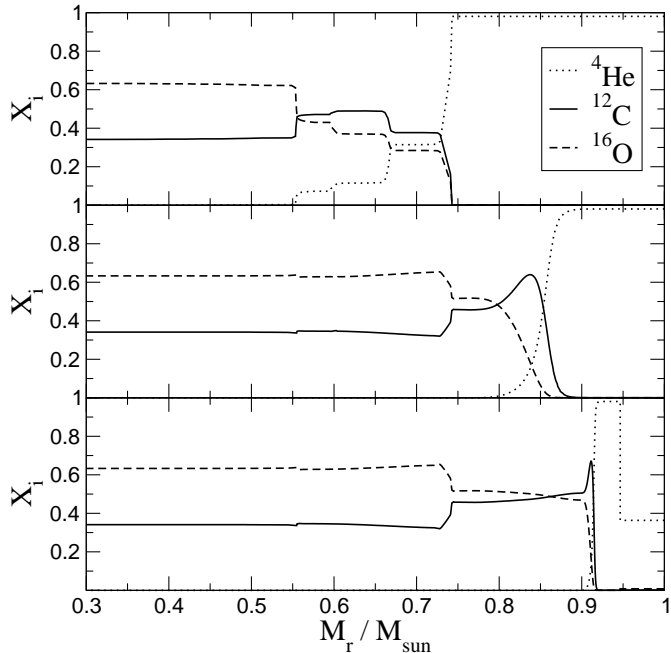}
\caption{Same as figure 2 but for the  6-M$_{\odot}$ sequence (OV). Core
overshooting leads to a larger $^{12}$C/$^{16}$O  core by the
end of  central helium burning.}
\end{figure}

The inner  chemistry variations  that take place  along the  red giant
branch and early AGB evolution is well documented in Figs. 2 and 3 for
sequence NOV and OV, respectively. Specifically, we show the evolution
of the internal  helium, carbon and oxygen distribution  as a function
of  mass for  the  evolutionary  stages following  the  end of  helium
burning in the  core.  We begin by examining  the results for sequence
NOV depicted  in Fig. 2.  The  upper panel shows  the chemical profile
when  the helium  convective core  vanishes leaving  a  central oxygen
abundance  of 0.55 by  mass.  As  evolution proceeds,  the helium-rich
layers  overlying the  former convective  core are  radiatively burnt,
giving  rise  to  an  off-centred   peak  in  the  carbon  and  oxygen
abundances.   This is  shown  by the  middle  panel of  Fig. 2,  which
corresponds  to the moment  when the  star surface  luminosity exceeds
\all_lsun= 3.72 for  the first time after $ 4.25  \times 10^{7}$ yr of
evolution.  Because  the  oxygen  abundance, and  therefore  the  mean
molecular  weight, decreases inwards  at $M_r  \approx$ 0.45  \msun, a
salt finger instability characterized by a large diffusion coefficient
develops  at  this  point.    The  resulting  salt  finger  mixing  is
responsible for the  redistribution of the innermost $^{12}$C/$^{16}$O
profile,  redistribution  that takes  place  during  the following  $3
\times 10^{5}$ yr and which is  documented by the bottom panel of Fig.
2.  Note  also  that  during  this  time  interval  the  mass  of  the
$^{12}$C/$^{16}$O core has increased  considerably by virtue of helium
shell  burning, reaching  0.925  \msun  \ by  the  re-ignition of  the
hydrogen  shell before  the  occurrence of  the  first helium  thermal
pulse.  The  behaviour for sequence OV  is detailed in  Fig.  3. Here,
the size  of the $^{12}$C/$^{16}$O  core emerging from  the convective
helium  core burning  (upper panel)  becomes substantially  larger (we
remind the  reader that sequence OV  has a lower  initial stellar mass
than  sequence  NOV).  Thus,  a  smaller fraction  of  helium  remains
unburned and  so a shorter pre-AGB  phase is expected.  In fact, about
$4.2  \times 10^{5}$  yr are  needed to  evolve from  the end  of core
helium burning to the onset of the thermally pulsing AGB phase (bottom
panel), as compared with the $9.3
\times 10^{5}$ yr for sequence NOV. The carbon and oxygen distribution
in the core is clearly different from the case without overshooting, a
feature which  is expected  to bear its  signature in  the theoretical
period  spectrum of ZZ  Ceti stars  (see Section  4).  Note  also that
sequence  OV is  characterized  by a  somewhat  larger central  oxygen
abundance (0.63 by  mass) than expected for sequence  NOV.  We want to
stress  again  that  breathing  pulses  have been  suppressed  in  our
calculations.   The  suppression  of  breathing  pulses  inhibits  the
formation  of  $^{12}$C/$^{16}$O   cores  with  large  central  oxygen
abundances  (see  Straniero  et  al.  2003  and  references  therein).
However, when central helium has been substantially depleted, sequence
OV  experiences a  small growth  of the  convective helium  core, 
which increases the central
helium abundance from 0.07 to 0.10. This fact is, in part, responsible
for the  higher central  oxygen abundance that  characterizes sequence
OV.

\begin{figure}
\centering
\includegraphics[clip,width=250pt]{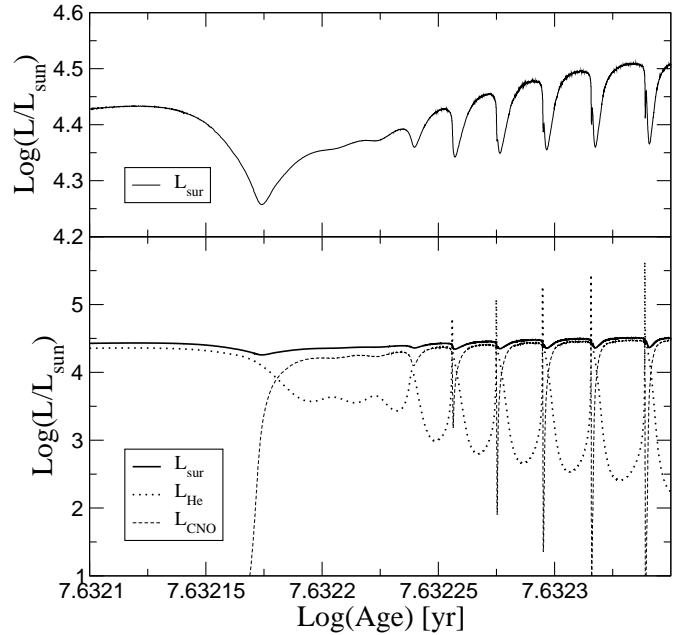}
\caption{Upper panel: the time-dependence of surface luminosity (in 
solar units)  for the 7.5-M$_{\odot}$ sequence (NOV)  during the first
thermal  pulses at  the  AGB.  Bottom  panel:  the time-dependence  of
surface luminosity,  hydrogen- (as given  by CNO cycle  reactions) and
helium-burning luminosities (dashed and dotted lines respectively) for
the same  sequence as above.  The time-scale is  given in yr  from the
main sequence. A  total of 7 thermal pulses  have been computed before
the  WD progenitor  departs from  the  AGB as  a result  of mass  loss
episodes.}
\end{figure}

Towards the  end of the early  AGB phase, hydrogen is  re-ignited in a
thin shell and the star  begins to thermally pulse. Here, helium shell
burning  becomes unstable  (see Iben  \& Renzini  1983 for  a detailed
description of this phase). The time dependence of surface luminosity,
hydrogen- and helium-burning luminosities  for sequence NOV during the
first five thermal pulses is  given in Fig.  4.  The interpulse period
for this  sequence is  roughly $2.4\times 10^{3}$  yr.  Note  that the
helium  burning  rate  rises   very  steeply  at  each  pulse.   After
experiencing  7 thermal  pulses  and considerable  mass  loss, the  WD
progenitor  departs from  the AGB  and evolves  towards  the planetary
nebula region and eventually to the WD state.  In our simulation, when
departure from the AGB  occurs, stationary helium-shell burning mostly
contributes  to  the  star  luminosity, but  shortly  after,  hydrogen
burning  takes over.  During  the thermal  pulses, mass  loss episodes
reduce the stellar mass from 6.15 to 0.936
\msun. It is worth noting that during the interpulses the temperature
at  the  base of  the  convective  envelope  becomes high  enough  for
hydrogen-burning  reactions to  occur. The  nuclear processing  at the
base  of the  convective envelope  is  commonly referred  to as  ``hot
bottom burning'', a process  whose occurrence is strongly dependent on
the treatment  of convection.   As we mentioned,  we will  not discuss
this aspect  here (see D'Antona  \& Mazzitelli 1996 and  Mazzitelli et
al. 1999 for recent studies); suffice it to say that for sequence NOV,
temperatures at the bottom of  the convective envelope after the first
five pulses reach about $6 \times  10^{7}$ K, which is high enough for
an early,  albeit moderate, onset of  hot bottom burning  to occur. In
fact,  the surface carbon  abundance decreases  from 0.0021  to 0.0019
during this phase.
 
\begin{figure}
\centering
\includegraphics[clip,width=250pt]{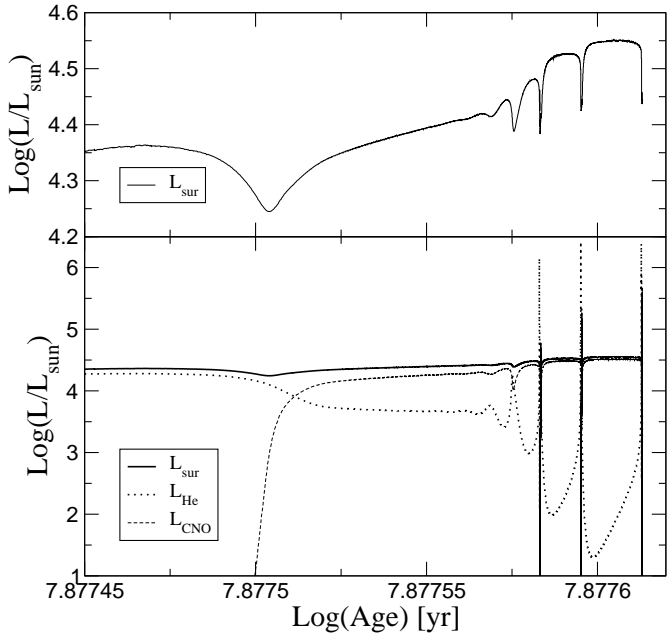}
\caption{Same as figure 4 but for the 6-M$_{\odot}$ sequence (OV). Here, 
the first three thermal pulses are shown.}
\end{figure}

\begin{figure}
\centering
\includegraphics[clip,width=250pt]{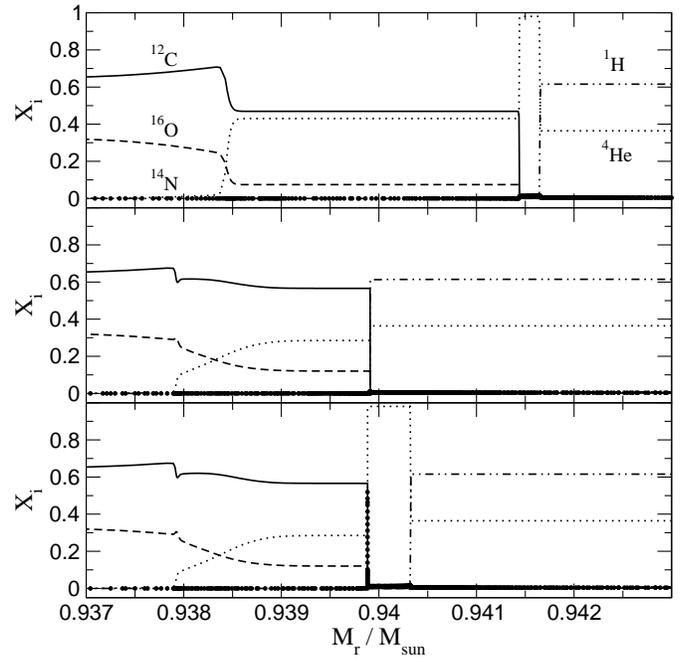}
\caption{Chemical profiles for the
6-M$_{\odot}$ sequence  (OV) at  three evolutionary stages  during the
2$^{\rm nd}$ thermal pulse.  At pulse peak (upper panel), the model is
characterized by a $^{4}$He-rich  buffer and an underlying region rich
in $^{4}$He,  $^{12}$C and $^{16}$O  (intershell).  The $^{4}$He-flash
convection zone and overshooting extending down to $M_r \approx$ 0.938
M$_{\odot}$ are  responsible for the redistribution  of the intershell
abundances  (middle panel).  Note  also that  envelope convection  has
penetrated  down  to  $M_r$=  0.9399 M$_{\odot}$,  sweeping  away  the
$^{4}$He   buffer   and    reaching   $^{12}$C-rich   regions   (third
dredge-up). During the interpulse, the helium buffer is built-up again
(bottom panel). Note also the  formation of a small $^{14}$N-pocket at
the base of the helium buffer.}
\end{figure}

The presence of overshooting from below the convective regions affects
the  evolution during  the thermally  pulsing AGB  phase.  The  run of
luminosities for  sequence OV  is detailed in  Fig.  5. Note  that the
trend of the  luminosity evolution is different from  sequence NOV. In
particular, the  rise in  the helium shell  burning luminosity  at the
peak of the  pulse is much more noticeable than  in sequence NOV. This
can be understood  on the basis that overshoot from  the bottom of the
helium-flash convection  zone carries some helium  from the intershell
region into deeper  layers, where it is burnt  at higher temperatures.
This picture can be visualized by  examining Fig.  6, in which we show
the internal abundance distribution  versus mass coordinate during the
second thermal pulse of  sequence OV. The region illustrated comprises
the  base of  the  hydrogen/helium envelope,  the  almost pure  helium
buffer,  the intershell region  and the  top of  the $^{12}$C/$^{16}$O
core. The upper panel corresponds to  the moment at which the model is
at  the  peak  of  its  second  thermal  pulse.  Here,  the  model  is
characterized  by  a  short-lived  (few years)  convection  zone,  the
product  of  the  huge  flux  of  energy  caused  by  unstable  helium
burning. This convection zone extends from $M_r \approx$ 0.9414 \msun\
down to the base of the helium-burning region at 0.9388 \msun.  At the
bottom of this helium-flash  convection zone there exists an overshoot
region  stretching  down to  the  underlying  carbon-rich layers  (the
$^{12}$C/$^{16}$O core)  at about 0.9383  \msun.  As a result  of this
overshoot region, larger amounts of carbon and oxygen are mixed up, as
compared with the situation  in which overshooting is neglected.  This
mixing  episode is  particularly important  regarding the  outer layer
chemical  stratification of  the emerging  WD remnant  (see  Herwig et
al.  1997).  As  can be  seen from  the middle  panel of  Fig.  6, the
resulting     abundances     of     the    intershell     zone     are
($^{4}$He/$^{12}$C/$^{16}$O)= (28/56/12)  as compared with (74/23/0.4)
for sequence NOV\footnote{The  quoted $^{4}$He/$^{12}$C values for the
intershell  abundances  change somewhat  with  subsequent pulses,  see
Herwig  (2000).}.    These  results  quantitatively   agree  with  the
predictions of Herwig (2000) for the case of lower stellar masses than
considered here.   Note also  that envelope convection  (and overshoot
from below)  has penetrated into  deeper layers down to  $M_r$= 0.9399
M$_{\odot}$,      reaching      $^{12}$C-rich      regions      (third
dredge-up\footnote{A process which is  also favoured by the occurrence
of  overshoot at the  bottom of  the helium-flash  convection zone.}).
During this process, the helium buffer, which was built-up by hydrogen
shell  burning  during  the   preceding  pulse,  is  completely  wiped
out. These episodes take place  while hydrogen shell burning is almost
extinct.  Sequence  OV experiences important dredge-up  of carbon even
at the  first pulse; however,  hot bottom burning taking  place during
the  following  quiescent  interpulse  phase  is  so  efficient  (with
temperatures  at the  base  of the  convective  envelope exceeding  $7
\times  10^{7}$  K)  in  converting  carbon  into  nitrogen  that  the
formation  of a  carbon star  is avoided  (at least  during  the first
thermal pulses). Interestingly,  about 25 \% of the  luminosity of the
star during this phase is produced within the convective envelope.

Bottom panel of Fig. 6 shows  that the helium buffer is built-up again
during   the   interpulse   phase   and   a   next   pulse   is   thus
initiated. Finally,  another feature predicted by  our calculations is
the presence of  a small radiative $^{14}$N-pocket at  the base of the
helium buffer (see  also Herwig 2000 for a  similar finding). In fact,
during  the  third  dredge-up  diffusive  overshoot  has  led  to  the
formation of  a small region in  which hydrogen from  the envelope and
carbon from  the intershell region coexist  in appreciable abundances.
When this region heats up enough to re-ignite hydrogen burning, one of
the main results is the formation of $^{14}$N with abundances reaching
about 0.5. However, the mass range over which the $^{14}$N-rich region
extends amounts only to $\approx 4 \times 10^{-9}$ \msun.  Eventually,
the  $^{14}$N-pocket is  swept into  the helium-flash  convection zone
during the next pulse.

\subsection{White dwarf evolution}

\begin{figure}
\centering
\includegraphics[clip,width=250pt]{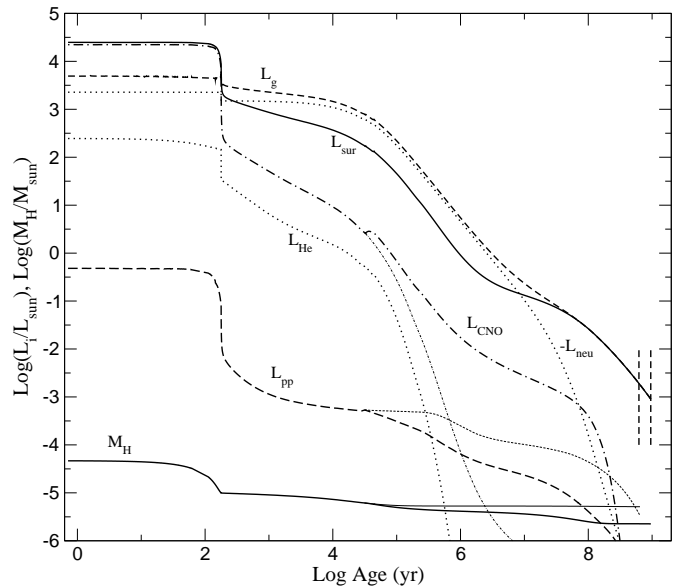}
\caption{Time-dependence of  different  luminosity
contributions and  mass of hydrogen  envelope ($M_{\rm H}$)  (in solar
units)  for  the   0.936-M$_{\odot}$  WD  remnant  of  7.5-M$_{\odot}$
evolution   (sequence  NOV):   surface   luminosity,  $L_{\rm   sur}$,
luminosity due to proton-proton reactions, $L_{\rm pp}$, CNO bi-cycle,
$L_{\rm CNO}$, helium burning,  $L_{\rm He}$, neutrino losses, $L_{\rm
neu}$ and  rate for  gravothermal (compressional plus  thermal) energy
release $L_{\rm  g}$.  Thin lines  depict $L_{\rm CNO}$,  $L_{\rm pp}$
and $M_{\rm H}$ for the  case when element diffusion is neglected. The
approximate  location of the  ZZ Ceti  instability strip  is indicated
with vertical dashed lines.  Note that  by the time the ZZ Ceti domain
is reached,  hydrogen burning is virtually extinguished.   As a result
of  hydrogen chemically  diffusing  inwards, nuclear  burning via  CNO
remains non-negligible  at late times and the  final hydrogen envelope
mass  becomes  a  factor  two  smaller  than  when  diffusion  is  not
considered.  Time is in years counted from the end of mass loss. }
\end{figure}

After considerable  mass loss,  the mass of  the hydrogen  envelope is
reduced so much that the  WD progenitor abandons the thermally pulsing
AGB phase and evolves towards large \teff\ values to becomes a WD.  By
the  end of mass  loss at  \teff\ $\approx  8000$ K,  the mass  of the
hydrogen envelope  of the 0.936-\msun\ remnant  (sequence NOV) amounts
to $M_{\rm H} \approx$ 5.5  $\times 10^{-5}$ \msun.  200 yr later, the
post-AGB remnant reaches  the point of maximum \teff\  and $M_{\rm H}$
is reduced  to $10^{-5}$  \msun\ as a  result of nuclear  burning.  We
mention that a similar value for  $M_{\rm H}$ has also been derived by
Bl\"ocker (1995) for his most massive WD remnant.  The time-dependence
of $M_{\rm  H}$ is displayed  in Fig. 7.  Notably, by the time  the ZZ
Ceti domain is reached, subsequent nuclear burning has further reduced
the hydrogen mass to $M_{\rm H} \approx$ 2.3 $\times 10^{-6}$
\msun.  This occurs as the WD evolves through the \teff\ (log$L$/L$_{\odot}$) 
range of 140000 : 25000 K (1.65 : $-$1.55).  Most of this reduction is
the result of nuclear burning via CNO of hydrogen chemically diffusing
inwards.   For  comparison, when  diffusion  is  neglected, the  final
hydrogen  content  remains  about  5.3 $\times  10^{-6}$  \msun\  once
nuclear burning via CNO  becomes virtually extinct at \teff\ $\approx$
50000 K after 1.6 $\times 10^{6}$ yr of evolution.  We want to mention
that because we have not  invoked additional mass loss episodes during
the planetary nebula stage or  early during the WD cooling branch, the
value of $M_{\rm H}$ should be considered as an upper limit. 
Because of the larger surface gravity characterizing massive
WDs, the resulting hydrogen envelope is less massive than in a typically 
0.6 \msun WDs. For instance, for a 0.563 \msun WD model, Althaus et al. 
(2002) find a hydrogen envelope mass of $7 \times 10^{-5}$ \msun.

In  Fig.   7  we also  show  as  a  function  of time  the  luminosity
contributions  due  to hydrogen  burning  via proton-proton  reactions
$L_{\rm pp}$  and CNO bi-cycle  $L_{\rm CNO}$, helium  burning $L_{\rm
He}$, neutrino losses $L_{\rm  neu}$, surface luminosity $L_{\rm sur}$
and gravothermal energy release  $L_{\rm g}$ for the 0.936-M$_{\odot}$
WD remnant (sequence NOV) from the  end of mass loss episodes near the
AGB to the domain of the ZZ Ceti stars. In Fig.  7 we also include the
predictions  for hydrogen burning  luminosities and  hydrogen envelope
mass  for the  situation  when element  diffusion  is neglected  (thin
lines).  Some features of this figure deserve comment.  In particular,
except  for  first  200  yr   when  the  remnant  evolves  to  its  WD
configuration, nuclear  burning never  constitutes the main  source of
surface  luminosity of  the star.   Note that  had diffusion  not been
considered, nuclear burning via  CNO cycle reactions would have ceased
after only $  10^{6}$ yr of evolution, a result  which is in agreement
with  that of  the Bl\"ocker  (1995) calculations.   By  contrast, CNO
reactions at  the base of the hydrogen  envelope remain non-negligible
for  a longer  period  of  time ($  2  \times 10^{8}$  yr)  in the  WD
evolution  if  diffusion  is  allowed  to operate.   This  is  because
chemical  diffusion  carries some  hydrogen  inwards  into the  helium
buffer and carbon upwards from the carbon-rich zone through the buffer
layer (see  later in this  section), thus favouring the  occurrence of
nuclear reactions.  However, {\it  by the time  the ZZ Ceti  domain is
reached, hydrogen burning becomes virtually extinct}.

From the  preceding section, we  saw that the  chemical stratification
for  both  the core  and  outer  layers of  the  WD  remnant is  quite
different  according  to whether  overshooting  is  considered or  not
during  the   pre-WD  evolution.   We  show   the  chemical  abundance
distribution at the start of the  WD cooling branch in the upper panel
of  Figs. 8  and 9  for the  0.936- and  0.94-M$_{\odot}$  WD remnants
(sequence NOV  and OV),  respectively.  The external  chemical profile
emerging  from  the  thermally  pulsing  AGB phase  (section  3.1)  is
characterized by a  region rich both in helium  and carbon (and oxygen
if overshooting  is considered), the relics of  the short-lived mixing
episode during  the peak of the last  helium pulse on the  AGB, and an
overlying helium-rich buffer. The mass of the helium buffer amounts to
$ 3 \times 10^{-4}$ and  $9.4 \times 10^{-4}$ \msun\ for sequences NOV
and OV, respectively. The buffer in sequence NOV is substantially less
massive  than in  sequence  OV  simply because  the  WD progenitor  in
sequence  NOV  departs from  the  AGB at  an  early  stage during  the
interpulse phase,  whilst the initial chemical  stratification for the
WD in sequence OV corresponds to a pre-WD structure at an evolutionary
stage well advanced in the interpulse phase.  Note also the signatures
left by  the third and last  dredge-up episode during  the last helium
thermal pulse in sequence OV,  in which convection penetrated into the
carbon-rich region  (see Fig. 6 for  details).  It is  also clear from
these figures  the markedly different chemical  profile resulting from
the consideration of core overshooting during central helium burning.

\begin{figure}
\centering
\includegraphics[clip,width=250pt]{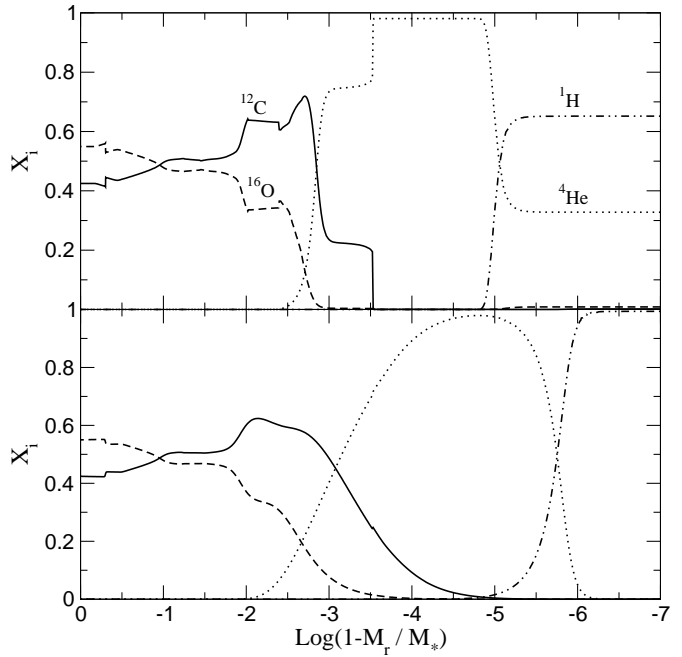}
\caption{Abundance by mass of $^{1}$H, $^{4}$He, $^{12}$C and
$^{16}$O  as   a  function  of   the  outer  mass  fraction   for  the
0.936-M$_{\odot}$  WD remnant  of 7.5-M$_{\odot}$  evolution (sequence
NOV) at the start of the  cooling branch (upper panel) and near the ZZ
Ceti instability  strip (bottom  panel).  Models are  characterized by
values (\all_lsun,  \lteff) of (2.25,  5.27) and (-2.92,  4.06) (upper
and bottom panel,  respectively).  Clearly, element diffusion strongly
affects the chemical profile.}
\end{figure}

\begin{figure}
\centering
\includegraphics[clip,width=250pt]{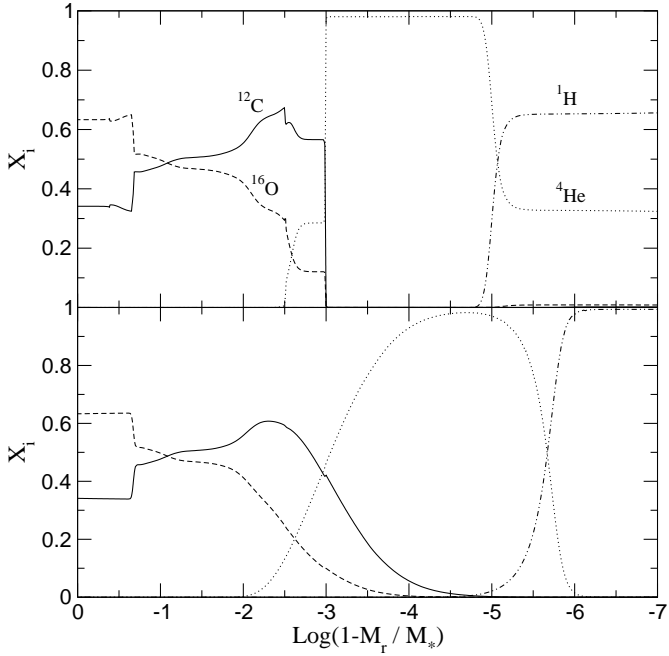}
\caption{Same as figure 8 but for the 0.94-M$_{\odot}$ WD remnant of 
6-M$_{\odot}$  evolution (sequence  OV).   Note that,  except for  the
inner  region of  the core,  element  diffusion acting  during the  WD
evolution is so efficient that the resulting abundance distribution at
the ZZ Ceti  stage does not depend on  whether overshooting during the
thermally pulsing AGB phase is considered or not.}
\end{figure}

\begin{figure*}
\centering
\includegraphics[clip,width=400pt]{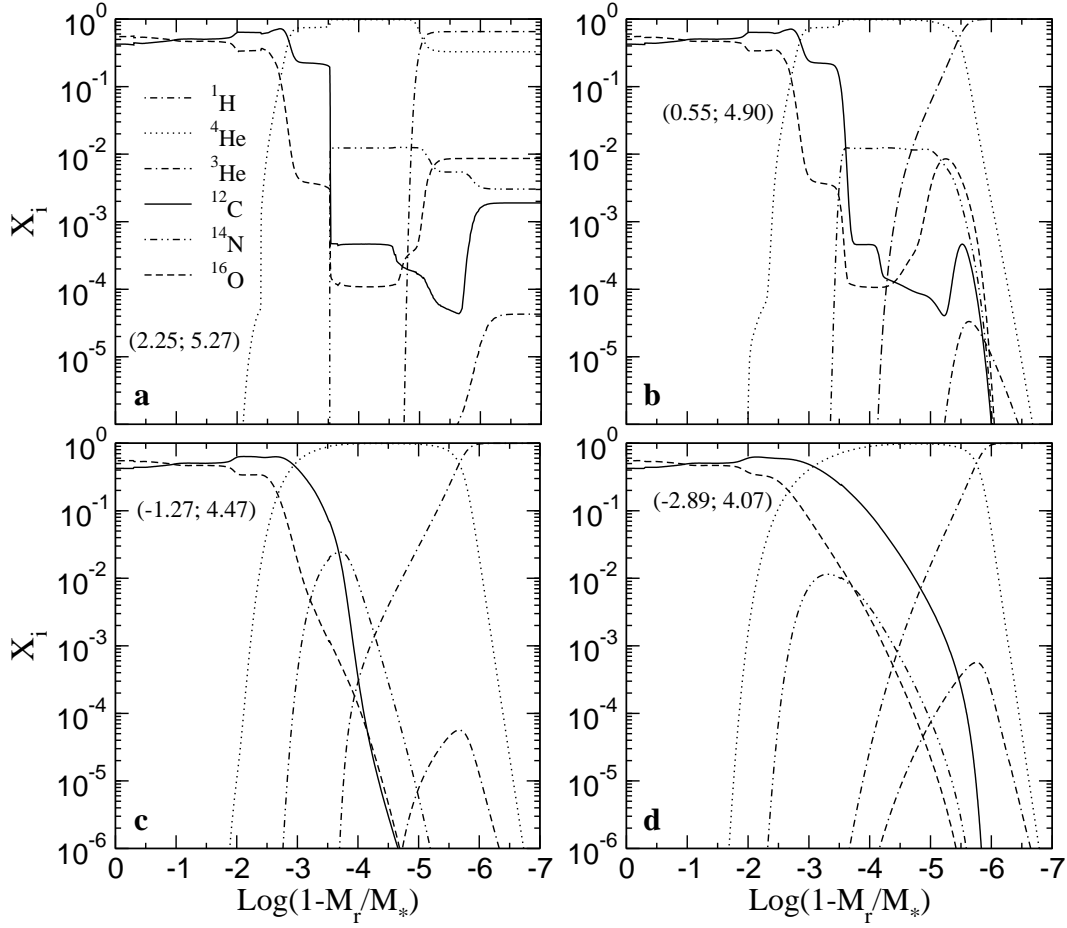}
\caption{Abundance  by mass  of $^1$H,  $^3$He, $^4$He,
$^{12}$C, $^{14}$N  and $^{16}$O in  terms of the outer  mass fraction
for  the  0.936-M$_{\odot}$ WD  remnant  of 7.5-M$_{\odot}$  evolution
(sequence NOV) at four  evolutionary stages characterized by values of
log$L$/L$_{\odot}$  and   log$T_{\rm  eff}$  (numbers   given  between
brackets).  Panel (a) corresponds to  the start of the cooling branch,
and panel (d)  to the ZZ Ceti domain.  At this  stage, note the inward
(outward)  extent of the  tail of  the hydrogen  (carbon) distribution
resulting from chemical diffusion.}
\end{figure*}

On the  cooling track,  the abundance distribution  of the  WD remnant
will  be  strongly modified  by  gravitational  settling and  chemical
diffusion.  This is  illustrated in the bottom panel of  Figs. 8 and 9
which show the chemical profiles at the ZZ Ceti stage for sequence NOV
and OV. These  figures emphasize the role of  element diffusion in the
external chemical  stratification of  massive ZZ Ceti  stars.  Indeed,
the shape  of the  chemical profile that  was built-up during  the AGB
phase is virtually  wiped out by diffusion processes  acting during WD
evolution.  In  particular, near- discontinuities left  by past mixing
episodes in the shape of  the external chemical interfaces (of primary
importance in  pulsation properties of WDs) are  strongly smoothed out
by the time the WD has approached  the hot edge of the ZZ Ceti domain.
It  is worth  noting that,  except  for the  inner part  of the  core,
element  diffusion  is  so  efficient  that  the  resulting  abundance
distribution  at  the  ZZ  Ceti  stage  does  not  depend  on  whether
overshooting during  the thermally pulsing AGB phase  is considered or
not. In  fact, the external chemical  profile at the ZZ  Ceti stage is
quite similar for both  sequences.  By contrast, towards the innermost
region, diffusion time scale becomes much longer than the evolutionary
time scale,  and the chemical  profile remains therefore  fixed during
the whole  WD evolution.\footnote{Except for the  minor mixing episode
at    the   innermost   $^{12}$C/$^{16}$O    core   induced    by   an
inward-decreasing   molecular  weight.}    Thus,   overshoot  episodes
occurring during  the pre-WD evolution leave  recognizable features in
the  chemical profile of  massive ZZ  Ceti stars  solely at  the inner
region of the core, features that are expected to leave their imprints
on the theoretical period spectrum  of these variable WDs (see section
4). As mentioned,  our model is characterized by  a chemical interface
in  which  helium,  carbon  and oxygen  in  non-negligible  abundances
coexist,  an interface  which,  at  the ZZ  Ceti  stage, has  extended
appreciably as a result of chemical diffusion.

The  evolution  of  the  chemical  abundance  distribution  caused  by
diffusion processes can best be seen in Fig. 10, which illustrates the
abundances by  mass of $^1$H,  $^3$He, $^4$He, $^{12}$C,  $^{14}$N and
$^{16}$O  for the  0.936-M$_{\odot}$ WD  remnant (sequence  NOV)  as a
function of the outer mass fraction at various epochs characterized by
values  of $\log{L/\rm  L_{\odot}}$ and  $\log{T_{\rm  eff}}$ (numbers
given   in   parentheses).   Panel   {\bf   a}   shows  the   chemical
stratification   at  the  start   of  the   cooling  branch   at  high
luminosities.    In  the   outermost   layers,  chemical   composition
corresponds basically  to that fixed  by the second  dredge-up episode
during the early AGB phase.   In the helium buffer, CNO abundances are
different from those in the  outer layers, because hydrogen burning in
earlier evolutionary phases processed  almost all the initial $^{12}$C
and  $^{16}$O  into  $^{14}$N.   Rapidly, gravitational  settling  and
chemical  diffusion  begin  to   play  an  increasing  role.   Indeed,
gravitational  settling causes hydrogen  to float  to the  surface and
heavier  elements to  sink down  to such  an extent  that  $4.2 \times
10^{5}$ yr  later and in the  absence of any  competing process, there
are no metals in the outermost $10^{-6}$ \msun\ of the star (see panel
{\bf b}).   With subsequent cooling, the effect  of chemical diffusion
is  clearly seen  at  the chemical  interfaces  where large  abundance
gradients exist.  As a result, in the helium buffer there is a tail of
hydrogen  reaching  hotter  layers  and  a tail  of  carbon  diffusing
outwards  from the  carbon-rich  zone (panel  {\bf  c}), thus  causing
hydrogen burning via CNO cycle to  be efficient for a longer period of
time than when diffusion is neglected (see Fig.  7). Notably, the tail
of  the  hydrogen  distribution   reaches  a  maximum  depth  by  this
epoch. Note also  that by this time, diffusion  has completely altered
the shape of the chemical profile below the helium buffer.  $7.3
\times  10^{8}$  yr  later,  the  star  reaches  the  ZZ  Ceti  domain
(panel {\bf d}). During this time, the tail of the carbon distribution
has diffused outwards until reaching  the outer boundary of the helium
buffer. Also,  the tail  of hydrogen begins  to retreat outwards  as a
result  of  gravitational  settling.   This  is  so  because  electron
degeneracy causes chemical diffusion  to become less important and the
inward diffusion  of hydrogen  is stopped.  It  is interesting  to see
that, despite the fact that  the diffusion of hydrogen and carbon into
the helium  buffer zone  has led to  sufficiently large  abundances of
these  two elements  by the  time the  ZZ Ceti  stage is  reached, the
temperature at the base of  the hydrogen envelope is too low ($\approx
5 \times  10^{6}$K) for hydrogen burning  to be of  any importance.  A
last remark  regarding the  influence of diffusion  is related  to the
evolution  of $^{12}$C and  $^{14}$N abundances  in the  buffer layer.
Indeed,  initially  the  $^{14}$N   abundance  in  the  helium  buffer
overwhelms that  of $^{12}$C  (see panel {\bf  a}), but before  the ZZ
Ceti  stage  is  reached,  $^{14}$N  is less  abundant  than  $^{12}$C
throughout   the  star  despite   nuclear  burning   having  processed
considerable  $^{12}$C into  $^{14}$N  along WD  evolution.  The  same
finding,  though to  a lesser  extent, has  been reported  by  Iben \&
MacDonald (1986) for the case of a 0.6 \msun WD.

In what follows we explore the implications for the global pulsational
properties of our models.

\section{Pulsational results}

For  the pulsational  analysis we  have  chosen two  template ZZ  Ceti
models with $T_{\rm eff}  \approx 12000$ K, corresponding to sequences
NOV and OV.  In addition,  we compare our results with the pulsational
predictions for  a 0.94-M$_{\odot}$, $T_{\rm eff}= 12000$  K WD model,
as given  by the stellar  modeling considered in Montgomery  \& Winget
(1999) (hereafter MW model).  We want to mention that despite the fact
that the  models are partially  crystallized (between $\sim  8\,$\% at
the blue  edge, to $\sim  25\,$\% at the  red edge of  the instability
strip), we have computed the theoretical period spectrum assuming that
the model interior  is in a completely uncrystallized,  fluid state. A
complete   discussion    of   the   pulsational    properties   taking
crystallization self-consistently into account would be quite lengthy,
so we have decided to postpone this to an upcoming communication,
for which we will  compute more massive WD models (and thus more suitable
for the  study of crystallized WDs) than attempted  here.

\subsection{The Ledoux term and the Brunt-V\"ais\"al\"a frequency}

The Brunt-V\"ais\"al\"a frequency ($N$),  a fundamental quantity in WD
pulsations,   is   computed by  employing   the   ``modified   Ledoux''
prescription. Specifically, $N$ is given by 
(see Brassard et al. 1991 for details)

\begin{equation} \label{ec:bv}
N^2   =   \frac{g^2\   \rho}{P}\   \frac{\chi_{_{\rm   T}}}{\chi_{\rho}}\
\left(\nabla_{\rm ad} - \nabla + B \right).
\end{equation}

The Ledoux term $B$, for the case of a multicomponent 
plasma (M-component plasma), is given by

\begin{equation}
B=-\frac{1}{\chi_{_{\rm  T}}}  \sum^{M-1}_{{\rm  i}=1}  \chi_{_{X_{\rm
i}}}
\frac{d\ln {X}_{\rm i}}{d\ln P}.
\end{equation}

Here, $\chi_{_{\rm T}}$ ($\chi_{\rm \rho}$) denotes the partial
logarithmic pressure derivative with respect to $T$ ($\rho$), $\nabla$
and  $\nabla_{\rm  ad}$  are  the  actual  and  adiabatic  temperature
gradients,  respectively, $X_{\rm  i}$  is the  abundance  by mass  of
species $i$, and

\begin{equation} 
\chi_{_{X_{\rm i}}}= \left( \frac{\partial \ln{P}}
{\partial \ln{X_{\rm i}}} \right)_{\rho,T,\{X_{\rm j \neq i}\} },
\end{equation}

\noindent where $P$ is the pressure. It is worth mentioning that the 
Ledoux term 
accounts explicitly  for the  contribution to $N$  from any  change in
composition in the model (the zones of chemical transition).  Brassard
et al.  (1992a) stress the relevance  of a correct treatment of $N$ at
the interfaces of chemical  composition transition zones in stratified
WDs, particularly in connection with the phenomenon of mode trapping.

\begin{figure} 
\centering
\includegraphics[clip,width=250pt]{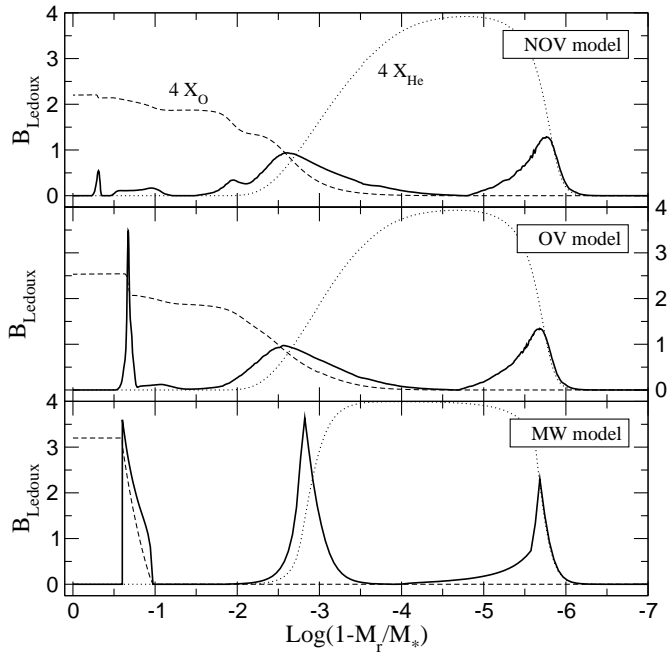}
\caption{The Ledoux term $B$ in terms of the outer mass fraction as 
predicted by the NOV, OV and MW models (upper, middle and bottom panel). 
In order to make  easier the identification of the various features 
exhibited by $B$, the mass fraction of $^{16}$O (dashed lines) and 
$^4$He (dotted lines) is depicted, with their magnitudes arbitrarily increased 
by a factor of 4 for clarity.}
\end{figure}

\begin{figure}
\centering
\includegraphics[clip,width=250pt]{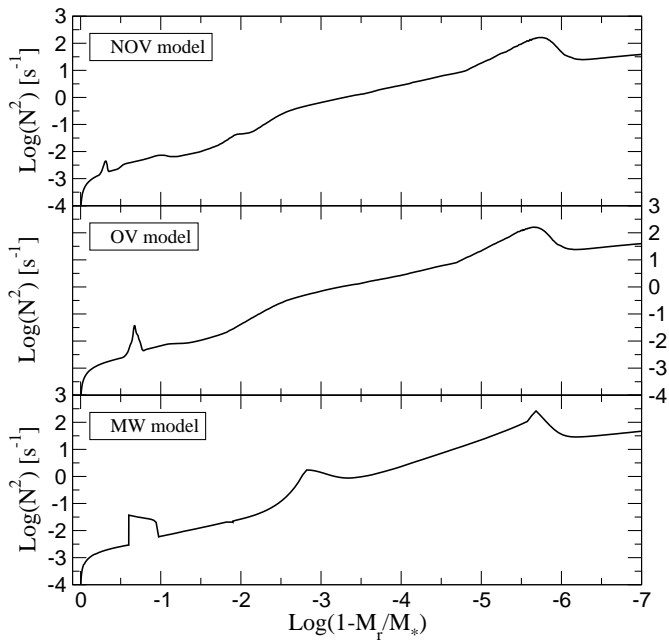}
\caption{Same as Fig. 11, but for the logarithm of the squared 
Brunt-V\"ais\"al\"a frequency.}
\end{figure}

In Fig. 11, the Ledoux term is plotted as a function of the outer mass
fraction for  NOV, OV and MW  models (upper, middle  and bottom panel,
respectively). In addition,  we have plotted the abundance  by mass of
$^{16}$O (dashed lines) and $^4$He (dotted lines) scaled by a factor 4
for a clear  and easy visual interpretation of  the different features
exhibited by $B$ (see from Eq.  19 the strong dependence of $B$ on the
slopes  of the  chemical abundances).  To  begin with,  note that  the
Ledoux term  in models NOV  and OV shows  smoother peaks in  the outer
layers,  as  compared with  the  case  of  MW model.   This  different
behaviour can  be in part understood  on the basis that  MW invoke the
diffusive equilibrium in the  trace element approximation (see Tassoul
et al. 1990)  to calculate the chemical profile  at the outer chemical
interfaces, an approximation that leads to pronounced peaks in the $B$
term.   Note that  the $B$  term for  OV model  is characterized  by a
high-amplitude peak  towards the innermost regions  where the presence
of overshooting has led to  a sharp variation of the $^{12}$C/$^{16}$O
profile (see Fig.  3 and 9). This behaviour is  in sharp contrast with
the situation  expected from  the case in  which core  overshooting is
neglected (upper panel of Fig. 11).  Note that the MW model also shows
a peak in  the core.  However, at variance with MW,  models NOV and OV
show an  extended bump  at the bottom  of the helium-rich  zone, which
again  is   a  result   of  the  broadness   and  smoothness   of  the
diffusion-modeled profiles.

In Fig. 12 we depict  the logarithm of the squared Brunt-V\"ais\"al\"a
frequency for the same cases  shown in Fig. 11. Each feature exhibited
by the Ledoux  term is clearly translated into  $N^2$.  Except for the
feature  induced  by  the  internal  oxygen-carbon  distribution,  the
Brunt-V\"ais\"al\"a frequency as predicted by the OV and NOV models is
very smooth. In particular,  note that the chemical transition regions
lead to smooth bumps at $\log q \sim -2.5$ and -5.5.  In contrast, the
Brunt-V\"ais\"al\"a  frequency  corresponding to  the  MW model  shows
enhanced peaks  at the  location of all  the chemical  interfaces (see
bottom panel of Fig. 12).

\subsection{Pulsational properties of the template models}

For the pulsation  analysis of the NOV and OV  models we have employed
the same  pulsational code  as in C\'orsico  et al. (2001;  2002).  We
refer the reader  to those papers and references  therein for details.
The boundary  conditions at the  stellar centre and surface  are those
given by Osaki  \& Hansen (1973) (see Unno et  al.  1989 for details).
Pulsation computations for the MW  model were carried out by employing
the   same    pulsational   code   as   in    Montgomery   \&   Winget
(1999)\footnote{This code  employ the outer boundary  conditions as in
Saio \& Cox (1980).}. For each computed mode we obtain the eigenperiod
$P_k$ (being  $k$ the radial  overtone of mode) and  the dimensionless
eigenfunctions  $y_1,\cdots, y_4$  (see Unno  et al.   1989  for their
definition).  Following   previous  studies  of   WD  pulsations,  the
normalization condition  adopted is $y_1=  1$ at the  stellar surface.
We also  compute the oscillation  kinetic energy, ($E_{\rm  kin}$; see
Eq. 1 of C\'orsico et al.   2002) and the weight function, $wf$, given
by  Kawaler et  al.  (1985).  The weight  function gives  the relative
contribution  of the  different  regions  in the  star  to the  period
formation (see  for details  Kawaler et al.  1985 and Brassard  et al.
1992ab).  Finally,  for each model  computed we derive  the asymptotic
spacing of periods as in Tassoul et al. (1990).

For our template models we have computed adiabatic g-modes with $\ell=
1, 2$  and $3$, with periods in  the range expected for  ZZ Ceti stars
(50 s $\lesssim P_k \lesssim$ 1300  s).  We begin by examining Fig. 13
which shows  the values  for the forward  period spacing  $\Delta P_k$
($\equiv P_{k+1}  - P_k$)  for $\ell= 1$  (upper panel) and  $\ell= 2$
(lower panel) in terms of  the periods computed.  An inspection of the
figure reveals that the amplitude of $\Delta P_k$ corresponding to the
NOV model are  typically lower as compared with the  results of the MW
model.   This   difference  is  understood  on  the   basis  that  the
Brunt-V\"ais\"al\"a frequency  of the NOV model is  smoother than that
of the  MW model  (compare upper  and bottom panel  of Fig.  12).  The
marked smoothness  of the $\Delta  P_k$ distribution in the  NOV model
implies that mode trapping is appreciably diminished in this model.

\begin{figure} 
\centering
\includegraphics[clip,width=250pt]{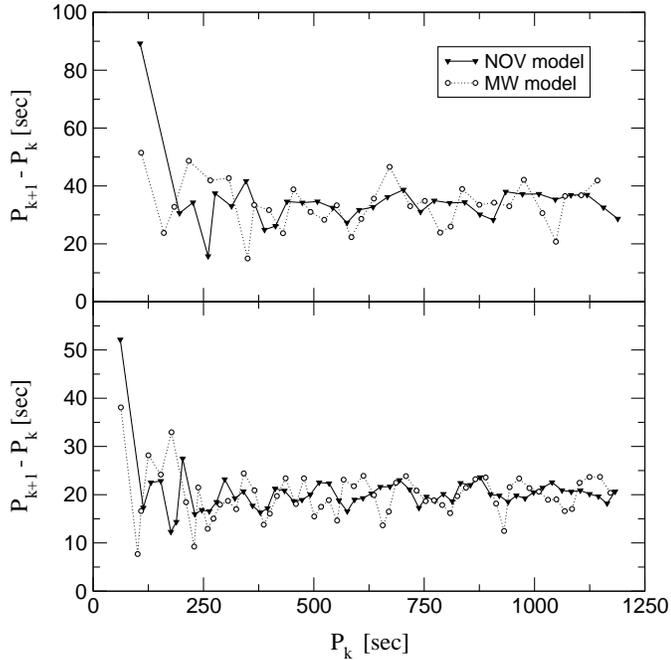}
\caption{Period spacing values for $\ell=  1$ (upper panel) and 2 
(bottom panel) in terms of the  computed periods, $P_k$ for the NOV 
model and MW predictions. Filled triangles and solid lines (empty dots 
and dotted lines) correspond to the NOV (MW) model.}
\end{figure}

In Fig. 14 we show the  same $\Delta P_k$-$P_k$ diagram as in Fig. 13,
but this time we compare the results  of the OV model with the MW one.
At first glance, the $\Delta  P_k$ values for both set of computations
exhibit similar amplitudes.  However,  a closer inspection of the plot
reveals   an  interesting   feature.   In   fact,  the   $\Delta  P_k$
distribution  corresponding to  the  OV model  clearly shows  periodic
minima with values  between minima almost constant and  tending to the
asymptotic value  (not plotted).  As we  shall see towards  the end of
this  section, the  minima in  $\Delta P_k$  correspond to  modes {\it
partially confined} to the deepest regions of the OV model as a result
of the  pronounced $N^2$-peak showed by  middle panel of  Fig. 12.  By
contrast, the $\Delta  P_k$ distribution for the MW  model shows clear
signals of mode trapping with  different amplitudes caused by the mode
trapping due to the presence  of {\it all} chemical interfaces in this
model.

\begin{figure} 
\centering
\includegraphics[clip,width=250pt]{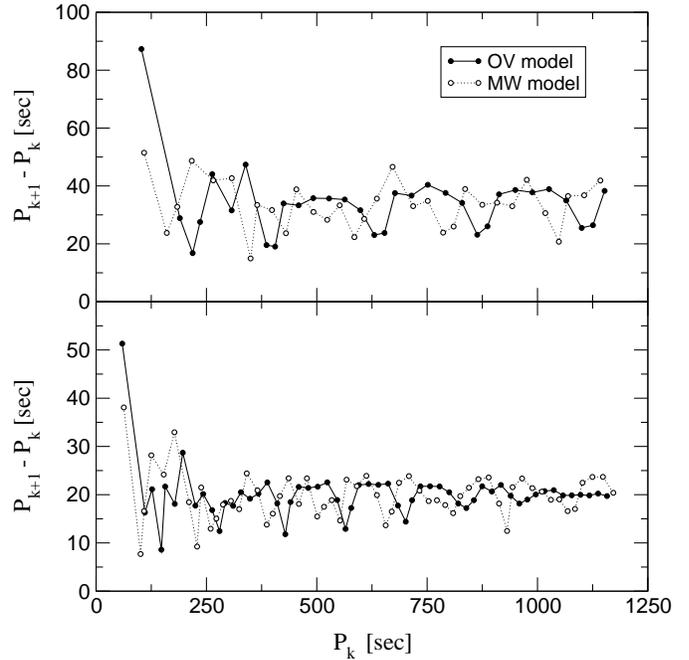}
\caption{Same as Fig. 13, but for OV and MW models. 
Filled dots and solid lines (empty dots 
and dotted lines) correspond to the OV (MW) model.}
\end{figure}

Finally,  in  Fig.  15   we  compare  the  $\Delta  P_k$  distribution
corresponding to  the NOV and OV  models. We can  clearly observe that
the non-uniformity in the period  spacing is much more apparent in the
OV model, as  expected from the shape of  $N^2$ corresponding to these
cases  (Fig. 12). As  stated before,  the minima  in the  $\Delta P_k$
distribution  for the  OV model  are associated  with  modes partially
confined to the high-density core  region placed between the centre of
the  model and  $\log q  \sim  -0.7$.  Consequently,  these modes  are
energy-enhanced,  so they  must exhibit  maxima in  the  $E_{\rm kin}$
distribution. To demonstrate this, we  show in Fig. 16 the oscillation
kinetic energy distribution corresponding to the OV WD model, where we
have labeled in particular the  $\ell= 2$ modes with overtones $k= 17,
19$ and  21. Note from  Fig. 15 that  the $\ell=2$, $k=  19$ eigenmode
corresponds  to a  minimum  of $\Delta  P_k$,  but it  has actually  a
$E_{\rm kin}$ value near a local maximum (Fig. 16).

\begin{figure} 
\centering
\includegraphics[clip,width=250pt]{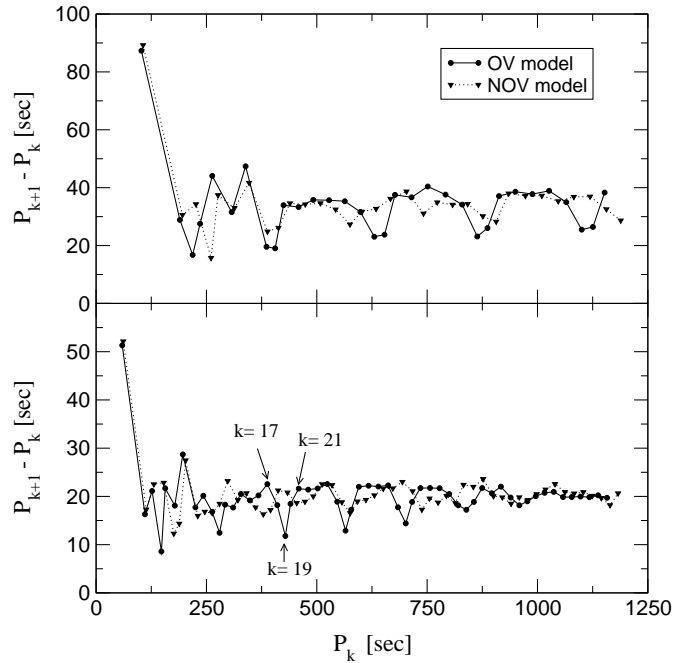}
\caption{Same as Fig. 12, but for OV and NOV models. 
Filled dots (triangles) correspond to the OV (NOV) model.}
\end{figure}

\begin{figure} 
\centering
\includegraphics[clip,width=250pt]{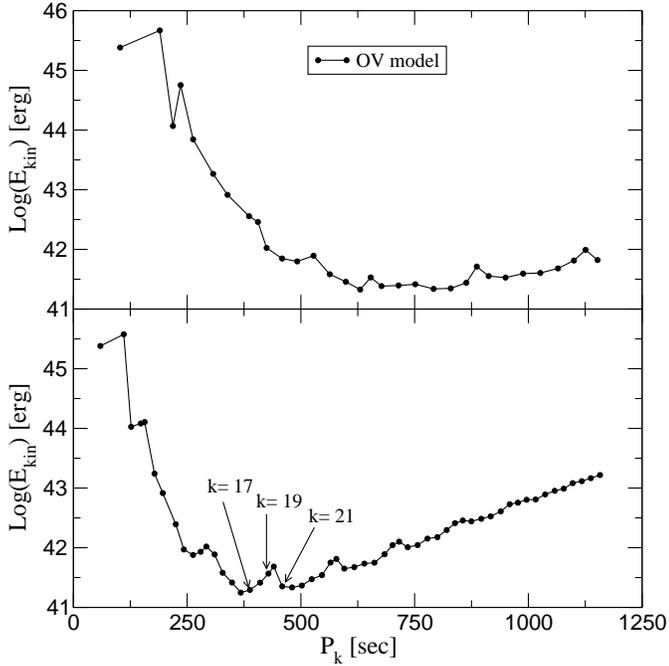}
\caption{Kinetic energy values for $\ell=  1$ and 2 in terms of the  
computed periods, $P_k$ for the OV model.}
\end{figure}

To  convince the  reader that  minima  in $\Delta  P_k$ correspond  to
energy-enhanced  modes,  we  depict   in  Fig.  17  the  amplitude  of
eigenfunctions $y_1$ and $y_2$ (corresponding to radial and tangential
displacements of matter, respectively) for  $\ell= 2$, $k= 17, 19$ and
21 modes at  the central region of the OV model.  In addition, we show
in  the  bottom  panel  the  weight function  corresponding  to  these
modes. An inspection of this  figure reveals that the amplitude of the
eigenfunctions as well as the weight function for the $k= 19$ mode are
noticeably  larger as  compared  with the  neighboring  ones ($k=  17,
21$). Note also the evident distortion  of the shape in $y_2$ and $wf$
for the  modes at the  location of the  acute peak in the  Ledoux term
($\log q \sim -0.7$; middle panel of Fig. 10).  This feature, which is
a direct consequence of the abrupt fall of $X_{^{16}{\rm O}}$, acts to
enlarge the  magnitude of $y_1$, $y_2$  and $wf$ of the  $k= 19$ mode,
but diminishes the  amplitude of these functions in  the case of modes
with $k= 17$ and 21.
 
\begin{figure}
\centering
\includegraphics[clip,width=250pt]{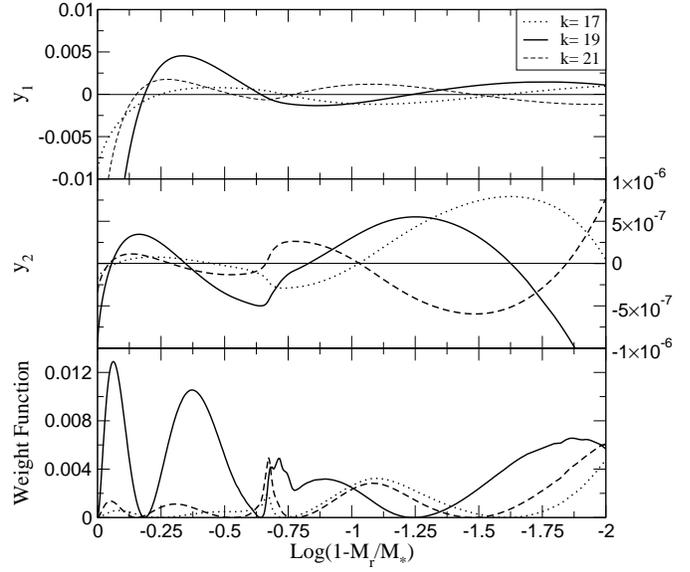}
\caption{Eigenfunctions $y_1$, $y_2$ and the weight function $wf$
(upper, middle and bottom panel, respectively) for modes with $\ell=  2$ 
and $k= 17$ (dotted lines), $k= 19$ (solid lines) and $k= 21$ 
(dashed lines). See text for details.}
\end{figure}

\subsection{Beyond the trace element approximation}

The trace element approximation has been a workhorse of chemical
profile calculations for many years. Sadly, as these and other
calculations have shown, it provides an inaccurate description of the
transition zone and its associated Ledoux term: it is too narrow, too
peaked, and its first derivative is not continuous.  In this section,
we present a new prescription which retains the simplicity of the
trace element approximation but which addresses its problems.

From  timescale  arguments  (e.g.,   Michaud  \&  Fontaine  1979)  and
numerical experiments,  we expect the  $^4$He/H transition zone  to be
near diffusive equilibrium,  so we are justified in  using Eq.~(A5) of
Arcoragi   \&   Fontaine   (1980)   for  the   equilibrium   profiles.
Traditionally,  the  trace  approximation  results from  solving  this
equation in the limit in which one of the two species is considered to
be a trace. However, assuming  complete ionization of both species, we
find that  Eq.~(A5) may still be  integrated in closed  form.  For the
$^4$He/H zone, we obtain

\begin{equation}
  \label{dif1}
  C\, \frac{P}{P_0} = \gamma^{\frac{1}{5}} (1+\gamma)^{\frac{18}{25}}
          (1+2 \gamma)^{-\frac{1}{10}} (1+6 \gamma)^{-\frac{1}{50}}, 
\end{equation}
where $\gamma=n_{\rm He}/n_{\rm H}$ is the ratio of number densities
of the two species, $P$ is the pressure, $P_0$ is some reference
pressure, and $C$ is a constant. Additionally, if we assume a
geometrically thin hydrogen layer (which is certainly valid), then the
pressure can be directly related to the mass depth, so that this
equation becomes
\begin{equation}
  \label{dif2}
  \tilde{C}\,\frac{q_s}{q_H} = \gamma^{\frac{1}{5}} (1+\gamma)^{\frac{18}{25}}
          (1+2 \gamma)^{-\frac{1}{10}} (1+6 \gamma)^{-\frac{1}{50}}, 
\end{equation}
where $q_s = 1-M_r/M_*$, $q_H = 1-M_H/M_*$, and $\tilde{C} \approx 1.764$.
Since the mass fraction of He is related to $\gamma$ by
\begin{equation}
  \label{dif3}
  X_{\rm He} = 
  \frac{A_{\rm He}\,n_{\rm He}}{A_{\rm H}\,n_{\rm H}+A_{\rm He}\,n_{\rm He}}
  = \frac{4 \gamma}{1+4 \gamma},
\end{equation}
Eq. (22) gives us a smooth relation for the mass fraction of He valid
from large abundances 
($X_{\rm He}\sim 1$, $\gamma \gg 1$) to low
abundances ($X_{\rm He}\ll 1$, $\gamma \ll 1$).

Since the treatment given above offers an easy implementation of
diffusion profiles, we would like to see how it compares with
self-consistently computed profiles. In Fig.~18 we compare the profile
calculated using Eq. (\ref{dif2}) with that obtained from an evolutionary
calculation assuming self-consistent, time-dependent diffusion.  We see
that our parameterized profile provides a good representation of the
overall profile shape, and that it reproduces the maximum height of
the Ledoux term to within $\sim\,$20\%. Since the numerical
implementation of this prescription is only slightly more expensive
computationally than that based on the trace element approximation, we
recommend Eq. (\ref{dif2}) as a reasonable replacement in codes which use a
pre-specified functional form for the profile shapes.

The main discrepancy between the two profiles shown in Fig.~18 is most
likely due to the neglect of electron degeneracy pressure which is
implicit in the use of Eq.~(A5) of Arcoragi \& Fontaine (1980).  Since
the deeper C-O/He transition zone is in a more degenerate environment,
as well as being much farther away from diffusive equilibrium, a
derivation along the lines which led to Eq. (\ref{dif2}) is more difficult to
justify; we therefore defer a discussion of a simple functional form
for its shape until a future paper. Finally, we note that the above
prescription for the $^4$He/H profile is aimed mainly at calculating
the Brunt-V\"ais\"al\"a frequency. As such, it may need to be
truncated at large depths in order to prevent unphysical situations
such as a thermonuclear runaway due to H burning.

\begin{figure}[t!]
\includegraphics[clip,width=250pt]{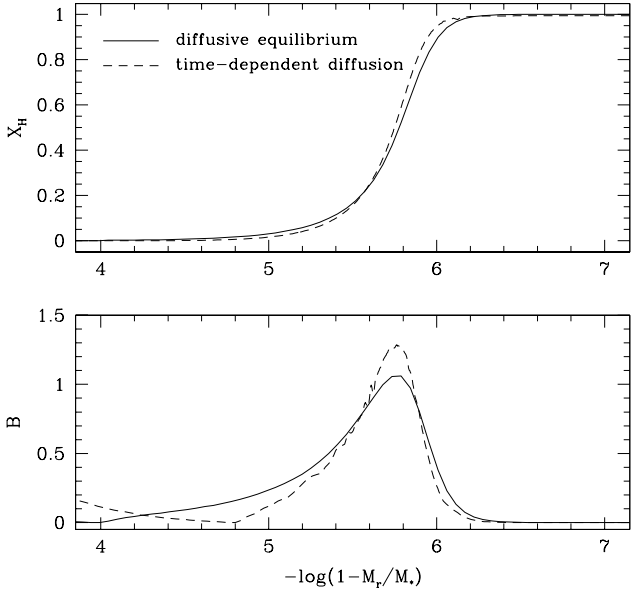}
  \caption{A comparison of the hydrogen mass fraction (upper panel)
    and the Ledoux $B$ term (lower panel) for the profile computed
    using Eq. (\ref{dif2}) (solid line), which assumes diffusive equilibrium,
    complete ionization, and an ideal gas equation of state, with that
    assuming full time-dependent diffusion (dotted line).}
  \label{comp}
\end{figure}

\section{Conclusions}

In this work we have computed new and improved evolutionary models for
carbon-oxygen DA WD stars appropriate for the study of massive ZZ Ceti
stars.  In  addition,  the  implications  of our  new  models  for  the
pulsational properties of massive ZZ  Ceti stars have been explored.
   To this  end,  we  have  followed the  {\it  complete}
evolution of massive WD progenitors from the zero-age
main sequence  through the thermally  pulsing and mass loss  phases on
the AGB to  the WD regime. Attention has been  focused on the modeling
of the chemical abundance distribution. In this regard, we developed a
time-dependent  scheme  for  the  simultaneous treatment  of  chemical
changes caused  by nuclear burning  and mixing processes.  Salt finger
mixing, semiconvection and diffusive  overshooting above and below any
formally  convective zone  have been  fully accounted  for  during the
pre-WD regime.  In this work, we  have taken into  account an extended
mixing  length theory  for fluids  with composition  gradients.  Also,
time-dependent  element diffusion  for multicomponent  gases  has been
considered during the WD evolution.

An important aspect of the  study has been to explore the implications
of  the occurrence of overshooting during the pre-WD evolution for  
the pulsational  properties of  massive  ZZ Ceti
stars.  To  this end,  we restrict ourselves  to examining two  cases of
evolution for the progenitor: sequence NOV based on the evolution of a
7.5-\msun \ initial  mass star  in which  overshooting  was not
considered and sequence OV based on the evolution of a 6-\msun \ star
with overshooting. For both  sequences, the mass of the resulting
$^{12}$C/$^{16}$O core is quite similar ($\approx 0.94$ \msun). This 
has allowed us in principle to compare the pulsational properties
of  carbon-oxygen massive ZZ  Ceti stars  {\it having the  same stellar
mass  but  being the  result  of  the  evolution of  progenitor stars  with
markedly   different  initial   masses.}
 
As for the main evolutionary results for the WD progenitor, we mention:

\begin{itemize}

\item Both of the sequences experience the second dredge-up episode during
which the surface composition is appreciably modified from the assumed 
interstellar abundances.

\item A salt finger instability develops after the end of
central helium burning, which is particularly noticeable in the case without 
core overshooting. This instability is responsible for the redistribution 
of the innermost $^{12}$C/$^{16}$O profile occurring before the end of
the early AGB phase.

\item Sequence OV  exhibits a sharp variation of the
$^{12}$C/$^{16}$O composition around $M_r \approx 0.72$ \msun, variation 
which leaves its signature on the theoretical period spectrum (see below).

\item During the thermally pulsing phase, the temperature at the base of the
convective envelope becomes high enough for hot bottom burning to occur. This 
is particularly noticeable for sequence OV.

\item The presence of overshooting below the bottom of the 
helium-flash convection
zone gives rise to an enhancement of oxygen and carbon in the intershell region
below the helium buffer, as compared with models without overshoot. This is in
agreement with Herwig (2000) for the case of low- and intermediate- mass 
stars. 

\item Sequence OV experiences the third dredge-up. At the end of 
this episode, a 
$^{14}$N-pocket is built-up at the base of the helium buffer. \\

For the WD evolution, our main conclusions are: \\

\item The mass of the hydrogen envelope left at the ZZ Ceti domain amounts
to $M_{\rm  H} \approx$  2.3 $\times 10^{-6}$ \msun.  This is about  half as
large as for the case when element diffusion is neglected.

\item As a result of chemical diffusion, CNO
reactions remain non-negligible
for a considerable  period of time ($ 2 \times  10^{8}$ yr) during the
WD evolution.   However, by  the time the  ZZ Ceti domain  is reached,
hydrogen burning becomes completely unimportant.

\item Element diffusion strongly modifies the shape of the chemical profile 
during WD  evolution. In particular, before the  ZZ Ceti stage is  reached, diffusion has already  smoothed out the chemical  profile to such  a degree that
the  resulting  external abundance distribution does not  depend  on  the
occurrence of overshoot episodes during the thermally pulsing phase on
the AGB. \\

Finally, the implications for the pulsational properties are: \\

\item Models with core overshooting (OV models) exhibit a much more
  pronounced non-uniformity in the spacing of consecutive periods, at
  variance with the situation without core overshooting.

\item OV models are characterized by the presence of modes which are
``core trapped'', i.e., modes with enhanced kinetic energy having
eigenfunctions with relatively large amplitudes at the innermost,
high-density regions of the models.

\item Pulsational properties are insensitive to the occurrence of overshoot
episodes during the thermally pulsing AGB phase. This is due to the
powerful action of element diffusion during WD evolution.

\item The period spectrum for our models is markedly different
from that of existing computations based on the use of the trace
element approximation for assessing the shape of the chemical interfaces.

\item The trace element approximation need not be made in the solution
  of the equation of diffusive equilibrium given in Arcoragi \&
  Fontaine (1980), which results in smooth profiles which are closer
  to the results of time-dependent diffusion.

\end{itemize}

We conclude that the pulsational  spectrum of massive ZZ Ceti stars as
predicted by our new and  improved evolutionary models turns out to be
substantially   different  from   that  of   previous   research.   In
particular,  because  element  diffusion  strongly  smoothes  out  the
chemical  profiles, the  mode trapping  caused by  the  outer chemical
interfaces  is  notably  diminished.   As a  result,  the  pulsational
properties  become  very  sensitive to  the shape  of the  {\it innermost}
chemical  profile, i.e., the  occurrence of  core overshoot  during the
evolutionary stages prior to the WD formation. This could in principle
be tested from observations of massive ZZ Ceti stars. We believe that
this aspect would deserve an exploration also in the frame
of less massive ZZ Ceti models than considered here.

In closing, we judge that the evolutionary modeling presented in this work
constitutes a physically sound and solid enough frame for exploring
the pulsational properties of crystallized ZZ Ceti stars. We will address 
this aspect in a future communication.

\begin{acknowledgements}

LGA warmly acknowledges T. Bl\"ocker for  sending us 
some reprints central to this work.
We also thank the suggestions and comments of our referee, D. Koester,
that strongly improved the original version of this work. This research was
supported by the Instituto de Astrof\'{\i}sica de La Plata and by the UK
Particle Physics and Astronomy Research Council. 

\end{acknowledgements}

\end{document}